\documentclass[longtitle,10pt]{elsarticle}
\usepackage{color}
\usepackage{amssymb}
\usepackage{amsfonts}
\usepackage{amsmath}
\usepackage{graphicx}
\usepackage{subfigure}

\begin{document}
\title{Building patterns by traveling dipoles and vortices in
two-dimensional periodic dissipative media}

\author[fr]{V. Besse\corref{cor1}}
\ead{valentin.besse@univ-angers.fr}

\author[fr]{H. Leblond}

\author[fr,horia,acad]{D. Mihalache}

\author[isr]{B.A. Malomed}

\cortext[cor1]{Corresponding author}

\address[fr]{LUNAM Universit\'{e}, Universit\'{e} d'Angers, Laboratoire de Photonique d'Angers, EA 4464, 2 Boulevard Lavoisier, 49000 Angers, France}
\address[horia]{Horia Hulubei National Institute for Physics and Nuclear Engineering, 30 Reactorului, Magurele-Bucharest, 077125, Romania}
\address[acad]{Academy of Romanian Scientists, 54 Splaiul Independentei, 050094 Bucharest, Romania}
\address[isr]{Department of Physical Electronics, Faculty of Engineering, Tel Aviv University, Tel Aviv 69978, Israel}



\begin{abstract}
We analyze pattern-formation scenarios in the two-dimensional (2D)
complex Ginzburg-Landau (CGL) equation with the cubic-quintic (CQ)
nonlinearity and a cellular potential. The equation models laser
cavities with built-in gratings, which stabilize 2D patterns. The
pattern-building process is initiated by kicking a compound mode, in
the form of a dipole, quadrupole, or vortex which is composed of
four local peaks. The hopping motion of the kicked mode through the
cellular structure leads to the generation of various extended
patterns pinned by the structure. In the ring-shaped system, the
persisting freely moving dipole hits the stationary pattern from the
opposite side, giving rise to several dynamical regimes, including
periodic elastic collisions, i.e., persistent cycles of elastic
collisions between the moving and quiescent dissipative solitons,
and transient regimes featuring several collisions which end up by
absorption of one soliton by the other. Another noteworthy result is
transformation of a strongly kicked unstable vortex into a stably
moving four-peaked cluster.
\end{abstract}



\maketitle

\section{Introduction}

The fundamental principle behind the creation of dissipative solitons is
that their stability relies upon the simultaneous balance of conservative
and dissipative ingredients in the underlying system \cite{DS}. These are
the diffraction and self-focusing nonlinearity in the conservative part of
the system, and linear and nonlinear loss and gain terms in the dissipative
part. Well-known physical realizations of such systems are offered by lasing
\cite{Rosa,lasers} and plasmonic \cite{plasmonics} cavities, the respective
models being based on the complex Ginzburg-Landau (CGL) equations with the
cubic-quintic (CQ) set of gain and loss terms, combined with the background
linear loss~\cite{lasers}. This combination is well known to maintain stable
localized modes \cite{Petv}. The CGL\ equations constitute a generic class
of dissipative pattern-formation models \cite{AK}, which find many other
applications, including bosonic condensates of quasi-particles in
solid-state media \cite{BEC}, reaction-diffusion systems \cite{MCCROSS}, and
superconductivity \cite{supercond}.

Originally, the CGL equation of the CQ type was introduced \cite{Petv} as a
model for the creation of stable two-dimensional (2D) localized modes.
Following this work, similar models were derived or proposed as
phenomenological ones in various settings. Many 1D and 2D localized states,
i.e., dissipative solitons, have been found as solutions of such equations
\cite{PhysicaD}-\cite{Vladimir}.

An essential ingredient of advanced laser cavities is a transverse periodic
grating, which can be fabricated by means of available technologies \cite%
{Jena}. In addition to the permanent gratings, virtual photonic lattices may
be induced in photorefractive crystals as interference patterns by pairs of
pump beams with the ordinary polarization, which illuminate the crystal
along axes $x$ and $y$, while the probe beam with the extraordinary
polarization is launched along $z$ \cite{Moti-general}. A 2D cavity model
with the grating was introduced in Ref. \cite{leblond1}. It is based on the
CQ-CGL equation including the cellular (lattice) potential, which represents
the grating. In fact, the laser cavity equipped with the grating may be
considered as a photonic crystal built in the active medium.
Periodic potentials also occur in models of passive optical systems, which
are driven by external beams and operate in the temporal domain, unlike the
active systems which act in the spatial domain \cite{tlidi94,Firth,advances}.

Localized vortices, alias vortex solitons, are an important species
of self-trapped modes in 2D settings. In uniform media, dissipative
vortex solitons cannot be stable without the presence of a diffusion
term, in the framework of the CGL equation (see, e.g., Ref.
\cite{Mihalache}). However, this term is absent in models of
waveguiding systems (it may sometimes be present in temporal-domain
optical models \cite{Fedorov}). Nevertheless, compound vortices,
built as complexes of four peaks pinned to the lattice potential,
may be stable in
models including the grating in the absence of the diffusion \cite{leblond1}%
. Using this possibility, stable 2D \cite{trapping_potentials_2D} and 3D
\cite{trapping_potentials_3D} vortical solitons have been found in the
framework of CGL equations including trapping potentials.

In a majority of previous works, the studies of various 2D localized
patterns have been focused on their stabilization by means of the
lattice potentials. Another relevant issue is mobility of 2D
dissipative solitons in the presence of the underlying lattice
(dissipative solitons may move without friction only if the
diffusion term is absent, therefore the mobility is a relevant issue
for the diffusion-free models of laser cavities). Localized modes
can be set in motion by the application of a kick to them, which, in
the context of the laser-cavity models, implies launching a tilted
beam into the system. Recently, the mobility of kicked 2D
fundamental solitons in the CQ-CGL equation with the cellular
potential was studied in Ref. \cite{cgl_mov}. It has been
demonstrated that the kicked soliton, hopping through the periodic
structure, leaves in its wake various patterns in the form of
single- or multi-peak states trapped by the periodic potential. In
the case of periodic boundary conditions (b.c.), which correspond to
an annular system, the free soliton completes the round trip and
hits the pattern that it has originally created. Depending on
parameters, the free soliton may be absorbed by the pinned mode
(immediately, or after several -- up to five -- cycles of
quasi-elastic collisions), or the result may be a regime of periodic
elastic collisions, which features periodic cycles of passage of the
moving soliton through the quiescent one.

A natural extension of the analysis performed in Ref. \cite{cgl_mov}
is the study of the mobility of kicked soliton complexes, such as
dipoles, quadrupoles, and compound vortices, and various scenarios
of the dynamical pattern formation initiated by such moving complex
modes, in the framework of the 2D CQ-CGL equation with the lattice
potential. This is the subject of the present work. In fact, such
configurations are truly two-dimensional ones, while the dynamical
regimes for kicked fundamental solitons, studied in Ref.
\cite{cgl_mov}, actually represent quasi-1D settings. The model is
formulated in Section II, which is followed by the presentation of
systematic numerical results for dipoles, quadrupoles, and vortices
of two types, onsite and offsite-centered ones (alias ``rhombuses"
and ``squares")  in Sections III, IV, and V, respectively. The paper
is concluded by Section VI.

An essential finding is that the interaction of a freely moving dipole with
pinned patterns, originally created by the same kicked dipole, gives rise to
new outcomes under the periodic b.c. In particular, the quiescent dipole can
be absorbed (\textquotedblleft cleared") by the moving one, which may have
obvious applications to the design of all-optical data-processing schemes,
where one may need to install or remove a blocking soliton.
Also noteworthy is the transformation of an unstable vortex by a strong kick
into a stable moving four-soliton cluster.

\section{The cubic-quintic complex Ginzburg-Landau model with the cellular potential}

The CQ-CGL equation with a periodic potential is written as

\begin{equation}
\frac{\partial u}{\partial Z}=\left[ -\delta +\frac{i}{2}\nabla _{\perp
}^{2}+(i+\epsilon )|u|^{2}-(i\nu +\mu )|u|^{4}+iV(X,Y)\right] u.
\label{CGL}
\end{equation}%
It describes the evolution of the amplitude of electromagnetic field $%
u(X,Y,Z)$ along propagation direction $Z$, with transverse Laplacian
$\nabla _{\perp }^{2}=\frac{\partial ^{2}}{\partial
X^{2}}+\frac{\partial }{\partial Y^{2}}$. Parameter $\delta $ is the
linear-loss coefficient, $\epsilon $ is the cubic gain, $\mu $ the
quintic loss, and $\nu $ the quintic self-defocusing coefficient (it
accounts for the saturation of the Kerr effect if $\nu >0$). The 2D
periodic potential with amplitude $V_{0}$ is taken in the usual
form, $V(X,Y)=V_{0}\left[ \cos (2X)+\cos (2Y)\right] $, where the
normalization of the field and coordinates is chosen so as to make
the normalized period equal to $\pi $, which is always possible. The
total power of the field is also defined as usual,%
\begin{equation}
P=\int \int \left\vert u\left( X,Y\right) \right\vert ^{2}dXdY.  \label{P}
\end{equation}

We solved CGL equation (\ref{CGL}) by means of the fourth-order
Runge-Kutta algorithm in the $Z$-direction, and five-point
finite-difference scheme for the computation of the transverse
Laplacian $\nabla _{\perp }^{2}$. Periodic boundary conditions
(b.c.) were used for the study of kicked dipoles and quadrupoles,
and absorbing b.c. for kicked vortices. In the latter case, the
absorbing b.c. was implemented by adding a surrounding
linear-absorption strip to the computation box. The absorption
coefficient varies quadratically with $X$ and $Y$ from zero at the
internal border of the strip to a value large enough to induce
complete absorption of any outgoing pulse, at its external border.
This smooth variation, if the width of the strip is not too small,
allows one to suppress any reflection from the absorption strip.

Values of coefficients chosen for numerical simulations are $\delta =0.4$, $%
\epsilon =1.85$, $\mu =1$, $\nu =0.1$, and $V_{0}=-1$. This choice
corresponds to a set of parameters for which the initial static
configurations for the dipoles, quadrupoles, and vortices are stable
(in-phase bound states of two dissipative solitons are possible too,
but, unlike the dipoles, with the phase shift of $\pi$ between the
bound solitons, they are unstable). The kick is applied to then in
the usual way, by adding the linear phase profile to
the initial field:%
\begin{equation}
u_{0}\left( X,Y\right) \rightarrow u_{0}\left( X,Y\right) \exp \left( i%
\mathbf{k}_{0}\cdot \mathbf{r}\right) ,  \label{kick}
\end{equation}%
where $\mathbf{r}\equiv \left\{ X,Y\right\} $. The key parameters are length
$k_{0}$ of kick vector $\mathbf{k}_{0}$, and angle $\theta $ which it makes
with the $X$-axis, i.e.,
\begin{equation}
\mathbf{k}_{0}=\left( k_{0}\cos \theta ,k_{0}\sin \theta \right) .
\label{theta}
\end{equation}

In the laser setup the kick corresponds to a small deviation of the
propagation direction of the beam from the $Z$ axis. If $K_{0}$ is the full
wave number and $\varphi $ is the deviation angle, the length of the
transverse wave vector in physical units is $K_{0}\sin \varphi $, which
corresponds to $k_{0}$ in the normalized form. Below, we investigate the
influence of kick parameters $k_{0}$ and $\theta $, defined as per Eq. (\ref%
{theta}), on a variety of multi-soliton complexes, which are created
by moving dipoles, quadrupoles, or vortices (of both onsite- and
offsite-centered types) in the 2D CGL medium with the cellular
potential.

\section{The pattern formation by kicked dipoles}

\subsection{Generation of multi-dipole patterns by a dipole moving in the transverse direction}

\begin{figure}[th]
\centering
\subfigure[]{\label{absdipoleinit}%
\includegraphics[width=5cm]{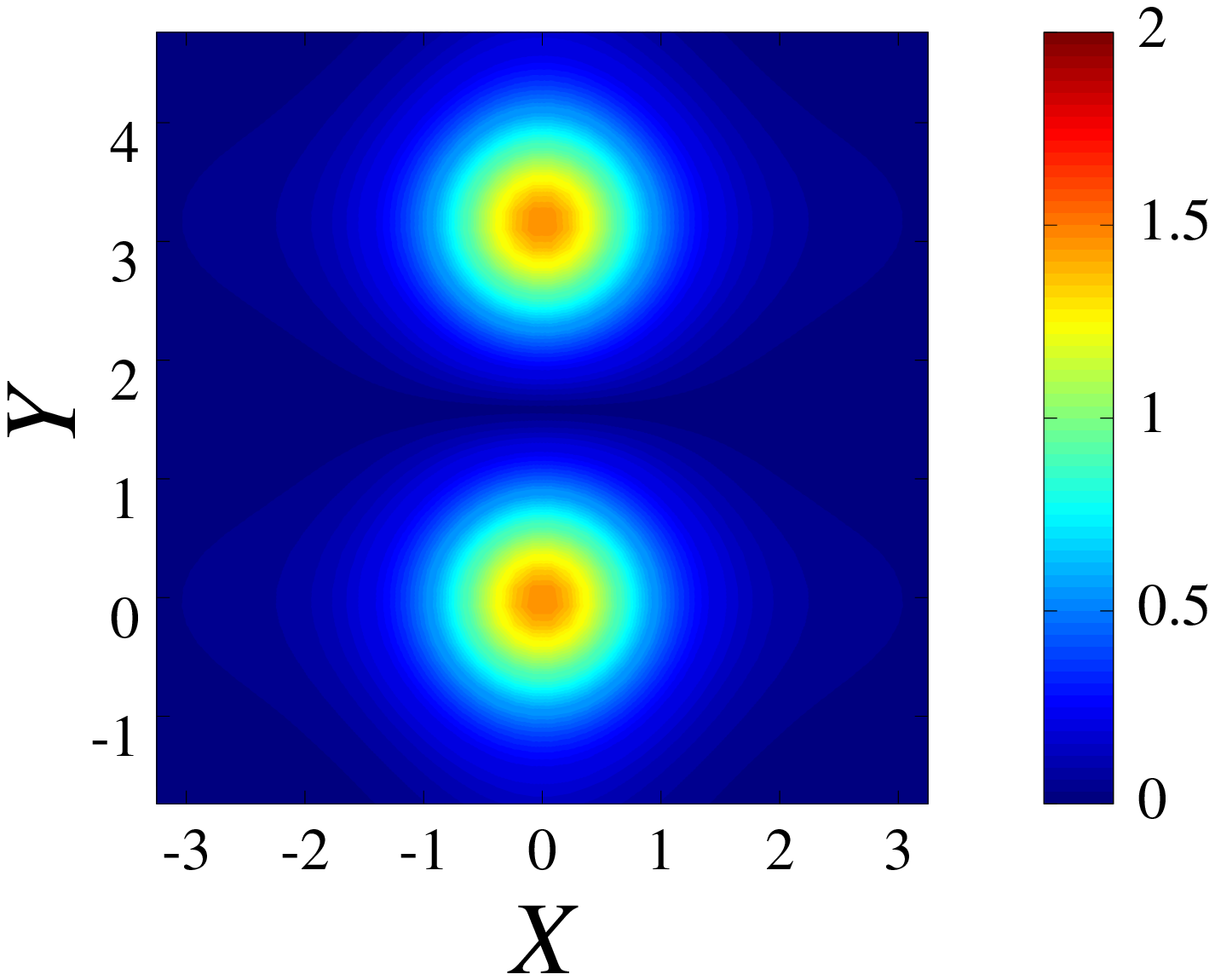}} \subfigure[]{%
\label{phasedipoleinit}\includegraphics[width=5cm]{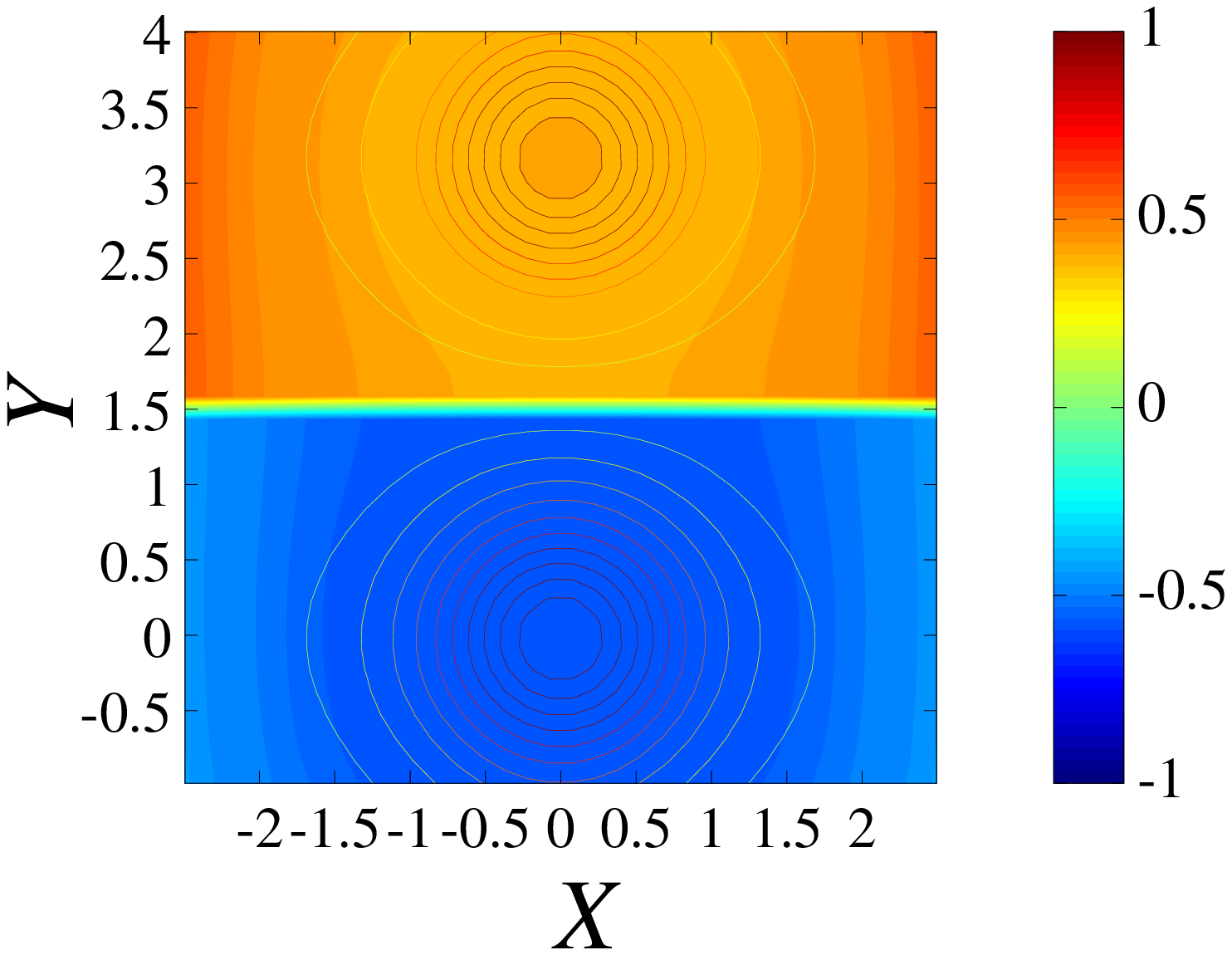}}
\caption{(Color online) The distribution of the amplitude (a) and phase (in
units of $\protect\pi $) (b) in the stable quiescent dipole mode.}
\label{dipoleinit}
\end{figure}

In this section we consider the simplest soliton complex in the form
of a stable vertical dipole, which consists of a pair of solitons
aligned along the $Y$-axis and mutually locked with phase difference
$\pi $, which is shown in Fig. \ref{dipoleinit}. The same color code
as in  Fig. \ref{absdipoleinit} is used in all figures showing
amplitude distributions throughout the paper. First, the dipole is
set in motion by the application of the kick in the horizontal ($X$)
direction (i.e.,
transversely to the dipole's axis), as per Eqs. (\ref{kick}) and (\ref{theta}%
) with $\theta =0$.

\begin{figure}[th]
\centering
\subfigure[]{\label{absdipolek016908theta0}%
\includegraphics[width=5cm]{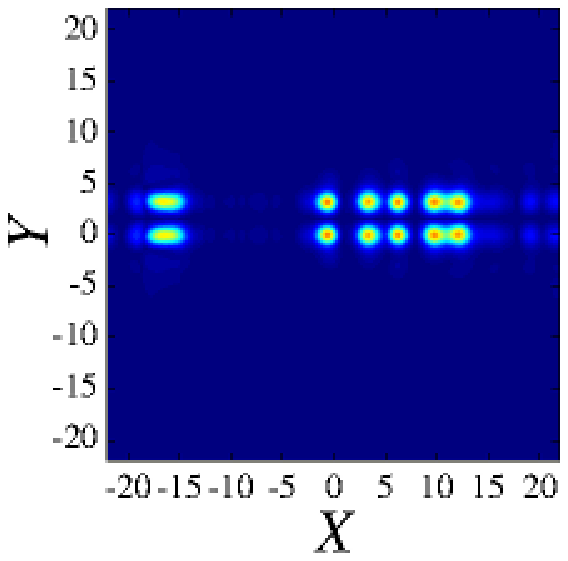}} %
\subfigure[]{\label{energdipolek016908theta0}%
\includegraphics[width=5cm]{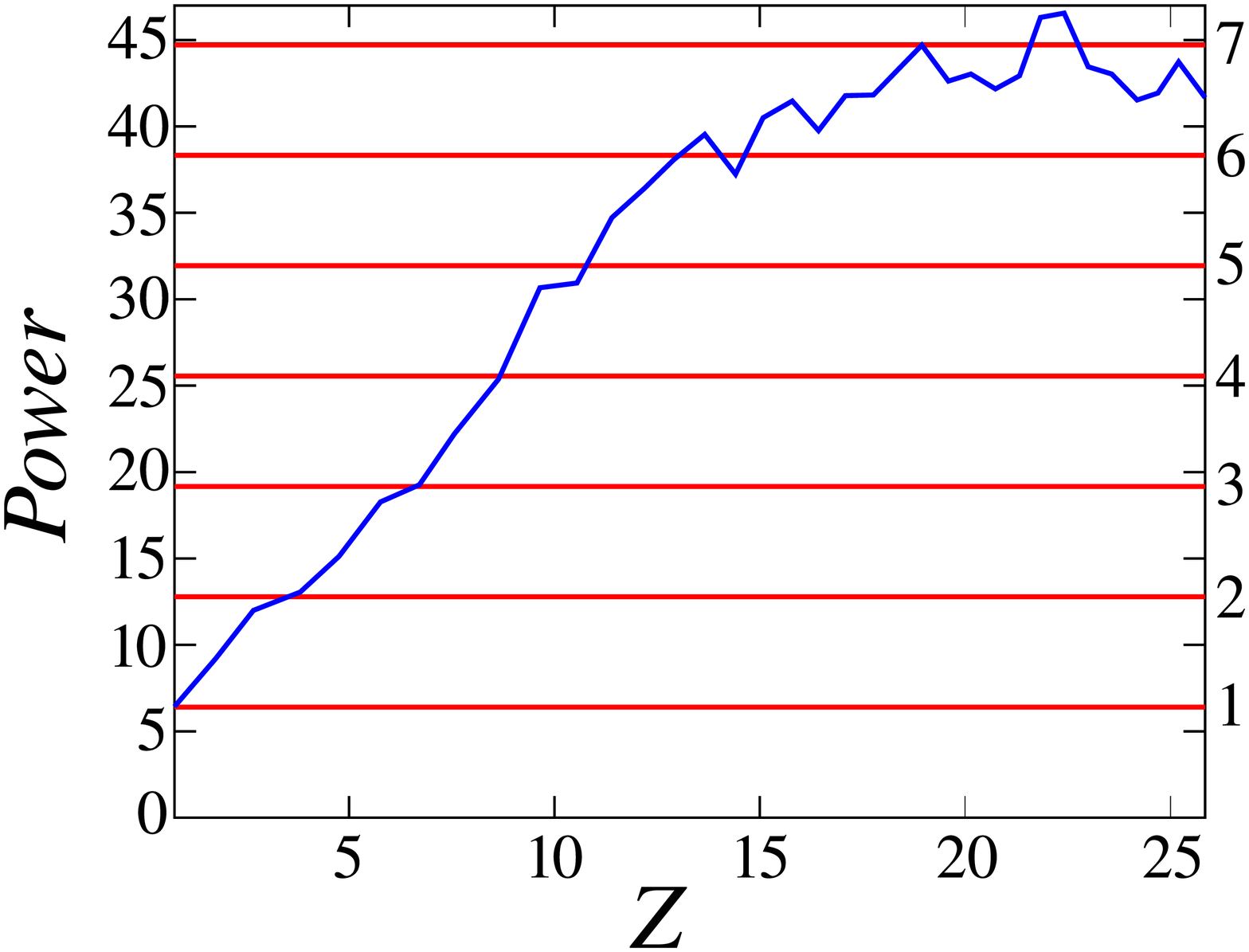}}
\caption{(Color online) (a) Field $\left\vert u(X,Y)\right\vert $ produced
by the horizontally kicked (with $\protect\theta =0$) vertical dipole at $%
Z=22.410$, for $k_{0}=1.665$. In this panel, the leftmost dipole is
moving to the right. The color code is the same as in Fig. 1a.  (b)
The evolution of the pattern produced by the horizontally kicked
dipole, shown in terms of the total power of the field as a function
of propagation distance $Z$. The set of horizontal red lines show
power levels corresponding to different numbers ($n$) of quiescent
dipoles.} \label{dipolek016908theta0}
\end{figure}

As shown in Fig. \ref{dipolek016908theta0}, the moving dipole
multiplies into a set of secondary ones, similar to the outcome of
the evolution of the kicked fundamental soliton \cite{cgl_mov}. Each
newly created dipole features the fixed phase shift $\pi $ between
two constituent solitons, and the entire pattern, established as the
result of the evolution, is robust. The particular configuration
displayed in Fig. \ref{dipolek016908theta0} is a chain of five
trapped dipoles, and a free one, which has wrapped up the motion and
reappears from the left edge, moving to the right, due to the
periodic b.c. Then, the free dipole will hit the pinned chain, and
will be absorbed by it, yielding a pattern built of five quiescent
dipoles. Immediately after the collision, the pattern features
intrinsic oscillations, which are gradually damped.

The snapshot shown in Fig. \ref{dipolek016908theta0} corroborates an
inference made from the analysis of numerical results: The largest number of
the dipoles generated by the initially kicked one is six, including one
moving dipole and five identical quiescent ones. It is worthy to note that,
as seen in Fig. \ref{energdipolek016908theta0}, in this case the total power
(\ref{P}) of the finally established set of six dipoles is close to the net
power corresponding to \emph{seven} quiescent ones, which is explained by
the observation that the power of the stably moving dipole is,
approximately, twice that of its quiescent counterpart.

To study the outcome of this dynamical pattern-formation scenario in a
systematic form, we monitored the number of output solitons as a function of
the kick's strength, $k_{0}$. These results are summarized in Fig. \ref%
{nombdipole}, which provides an adequate overall characterization of the
interactions, including a potential possibility to use these interactions
for the design of data-processing setups.

\begin{figure}[th]
\begin{center}
\includegraphics[width=10cm]{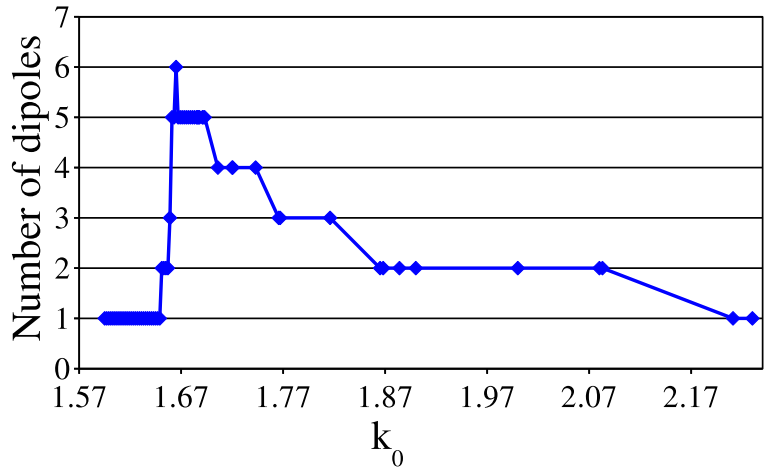}
\end{center}
\caption{(Color online) The number of dipoles in the final configuration
versus the kick strength, $k_{0}$, applied to the vertical dipole in the
horizontal direction.}
\label{nombdipole}
\end{figure}

Below the threshold value of the kick's strength, whose numerically found
value is
\begin{equation}
k_{0}^{(\mathrm{thr})}\left( \theta =0\right) \approx 1.651,  \label{thr0}
\end{equation}%
the kicked dipole exhibits damped oscillations, remaining trapped near a
local minimum of the cellular potential. Then, as seen in Fig. \ref%
{nombdipole}, the number of dipoles initially increases steeply with $k_{0}$%
, reaching (as mentioned above) a maximum of six at $k_{0}=1.665$.
It is
worthy to mention that this value is different from those, ranging in
interval $k_{0}\in \lbrack 1.6927,1.6942]$, in which the maximum number of
secondary solitons is reached in the case when the kick is applied to a
fundamental soliton \cite{cgl_mov}. This observation suggests that building
the structures by the kicked dipole does not merely reduce to the earlier
studied regime of the pattern formation by the individual solitons forming
the dipole. With the further increase of $k_{0}$, the number of solitons in
the output decreases by increasingly broad steps.

\subsection{Dynamical regimes initiated by the longitudinal kick applied to the dipole}

For the sake of the completeness of the description of the 2D system, we
have also simulated essentially quasi-1D dynamical regimes initiated by the
motion of the dipole kicked at angle of $\theta =\pi /2$, i.e., in the
longitudinal direction, see Eq. (\ref{theta}). This setting implies the
possibility to generate not only new dipoles but fundamental solitons as
well. It was found that the minimum value of the kick which is necessary to
set the dipole in motion is smaller in this case than the one given by Eq. (%
\ref{thr0}):
\begin{equation}
k_{0}^{(\mathrm{thr})}\left( \theta =\pi /2\right) \approx 1.303.
\label{thr_pi/2}
\end{equation}%
The results obtained for this configuration are summarized in Table \ref%
{table_dipole_theta0.5pi}. Above the threshold value (\ref{thr_pi/2}),
additional moving solitons are created: one at $k_{0}\in \left[ 1.304,1.875%
\right] $ and two in a narrow interval $k_{0}\in \left[ 1.880,1.885\right] $%
. Then, for $k_{0}\in \left[ 1.89,2.015\right] $, a new moving dipole
appears, which, as well as the original one, is oriented along the direction
of the motion, and accompanied by two moving solitons. For $k_{0}\in \left[
2.02,2.17\right] $, we have one moving soliton less, and at $k_{0}\in \left[
2.175,2.255\right] $ the original dipole disappears in the course of the
propagation, thus leaving one moving dipole and two moving solitons in the
system. At $k_{0}\in \left[ 2.26,2.36\right] $, we observe the same pattern
as for $k_{0}\in \left[ 2.02,2.17\right] $ (two dipoles and one moving
soliton). Then, for $k_{0}\in \left[ 2.365,2.46\right] $, the dipole splits
into two traveling solitons, with the upper one leaving a pinned soliton at
the site which it originally occupied. At higher values of the kick's
strength, the same pattern appears, except that the solitons do not leave
anything behind them, just traveling through the lattice.

\begin{table}[th]
\begin{center}
\begin{tabular}{|c||c|c|}
\hline
Behavior pattern & Range of $k_{0}$ & Number of new solitons \\
&  & along the $Y$-direction \\ \hline
1 dipole & $k_{0}\in \left[ 0,1.303\right] $ & 0 \\ \hline
1 dipole and 1 moving soliton & $k_{0}\in \left[ 1.304,1.875\right] $ & 1 \\
\hline
1 dipole and 2 moving solitons & $k_{0}\in \left[ 1.88,1.885\right] $ & 2 \\
\hline
2 dipoles and 2 moving solitons & $k_{0}\in \left[ 1.89,2.015\right] $ & 4
\\ \hline
2 dipoles and 1 moving soliton & $k_{0}\in \left[ 2.02,2.17\right] $ & 3 \\
\hline
1 dipole and 2 moving solitons & $k_{0}\in \left[ 2.175,2.255\right] $ & 2
\\ \hline
2 dipoles and 1 moving soliton & $k_{0}\in \left[ 2.26,2.36\right] $ & 3 \\
\hline
1pinned and 2 moving solitons & $k_{0}\in \left[ 2.365,2.46\right] $ & 2 \\
\hline
2 moving solitons & $k_{0}\in \left[ 2.465,\infty \right) $ & 0 \\ \hline
\end{tabular}%
\\[0pt]
\end{center}
\caption{The number of dipoles and fundamental solitons in the established
pattern versus the kick's strength $k_{0}$ directed along the dipole's axis (%
$\protect\theta =\protect\pi /2$). In the right column, a newly emerging
dipole (if any) is counted as two solitons.}
\label{table_dipole_theta0.5pi}
\end{table}

\subsection{Collision scenarios for moving dipoles in the system with periodic b.c.}

The above consideration was performed for a long system, before the
collision of the freely moving dipole with the static pattern left in its
wake, which should take place in the case of periodic b.c. In the
application to laser-cavity settings, the periodic b.c. in the direction of $%
X$ are relevant, corresponding to the cavity with the annular shape of its
cross section. The study of dynamical pattern-formation scenarios with the
periodic b.c. is also interesting in terms of the general analysis of models
based on the CGL equations \cite{cgl_mov}.

\begin{figure}[th]
\begin{center}
\subfigure[]{\label{absdipolek01865theta0XvsZ}%
\includegraphics[width=5cm]{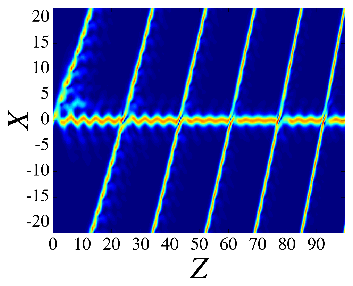}} %
\subfigure[]{\label{absdipolek01865theta0abs}%
\includegraphics[width=5cm]{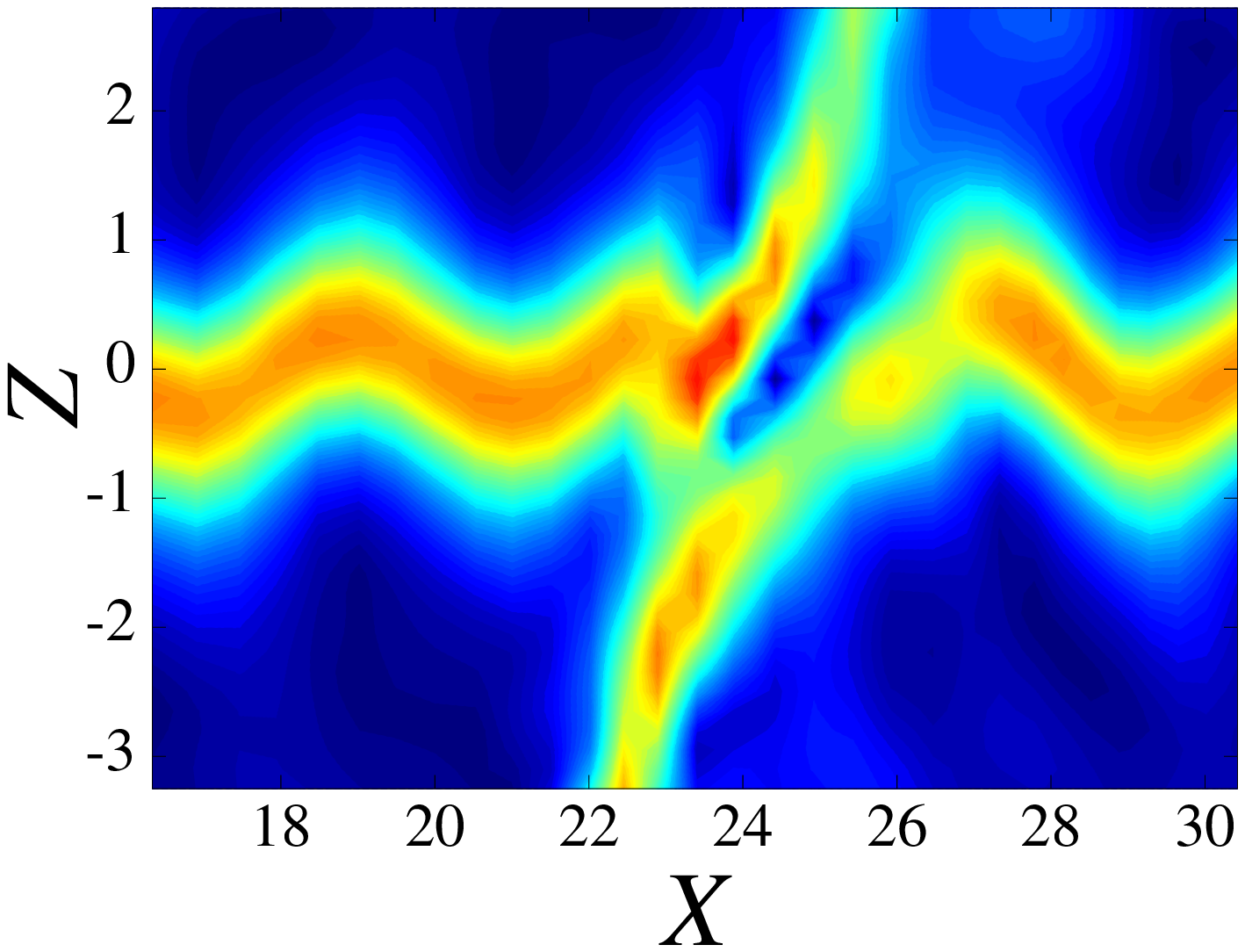}}
\end{center}
\caption{(Color online) (a) The cross section of field $\left\vert
u\left( X,Y,Z\right) \right\vert $ at $Y=0$, in the plane of $\left(
X,Z\right) $, for $k_{0}=1.865$. This is an example of the scenario
of the periodic elastic collisions, when the moving dipole repeats
elastic collisions with the quiescent one. (b) The close-up of the
elastic collision. The color code is the same as in Fig. 1a.}
\label{absdipolek01865theta0}
\end{figure}

Thus, under the periodic b.c., the freely moving dipole observed in Fig. \ref%
{dipolek016908theta0} will complete the round trip and will hit the trapped
chain of quiescent dipoles. Results of extensive simulations of this setting
are summarized in the list of three different outcomes of the collisions,
which feature persistent or transient dynamics (all the
regimes were observed for $\theta =0$, i.e., the transversely kicked dipole):

\begin{itemize}
\item[$\bullet $] The regime of the periodic elastic collisions,
corresponding to the periodically recurring passage of the moving dipole
through the quiescent one, see Fig. \ref{absdipolek01865theta0}. This
outcome takes place for $k_{0}\in \lbrack 1.865,1.868]$. Note that,
according to Fig. \ref{nombdipole}, in this region the pattern left in the
wake of the kicked dipole indeed amounts to another single quiescent dipole.

\item[$\bullet $] The transient regime, which features several quasi-elastic
collisions, before the moving dipole is eventually absorbed by the pinned
pattern, which is a bound complex of two dipoles, see Fig. \ref%
{absdipolek01816theta0}. This transient regime occurs around $k_{0}=1.816$,
in which case Fig. \ref{absdipolek01865theta0} confirms that the moving
dipole leaves a set of two additional dipoles in its wake.

\item[$\bullet $] The regime of \textquotedblleft clearing the obstacle",
opposite to the previous one: It features several elastic collisions, before
the pinned dipole is absorbed by the moving one, see Fig. \ref%
{absdipolek01884theta0}. This happens for $k_{0}\in \lbrack 1.884,1.9]$ and
around $k_{0}=2.083$ (in this region, Fig. \ref{nombdipole} confirms that
the moving dipole creates, originally, a single quiescent one).
\end{itemize}

\begin{figure}[th]
\begin{center}
\subfigure[]{\label{absdipolek01816theta0XvsZ}%
\includegraphics[width=5cm]{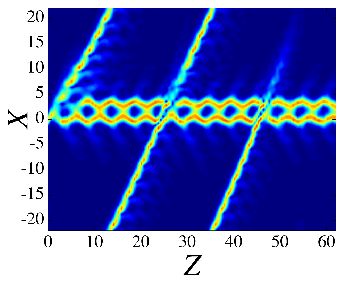}} %
\subfigure[]{\label{absdipolek01816theta0abs}%
\includegraphics[width=5cm]{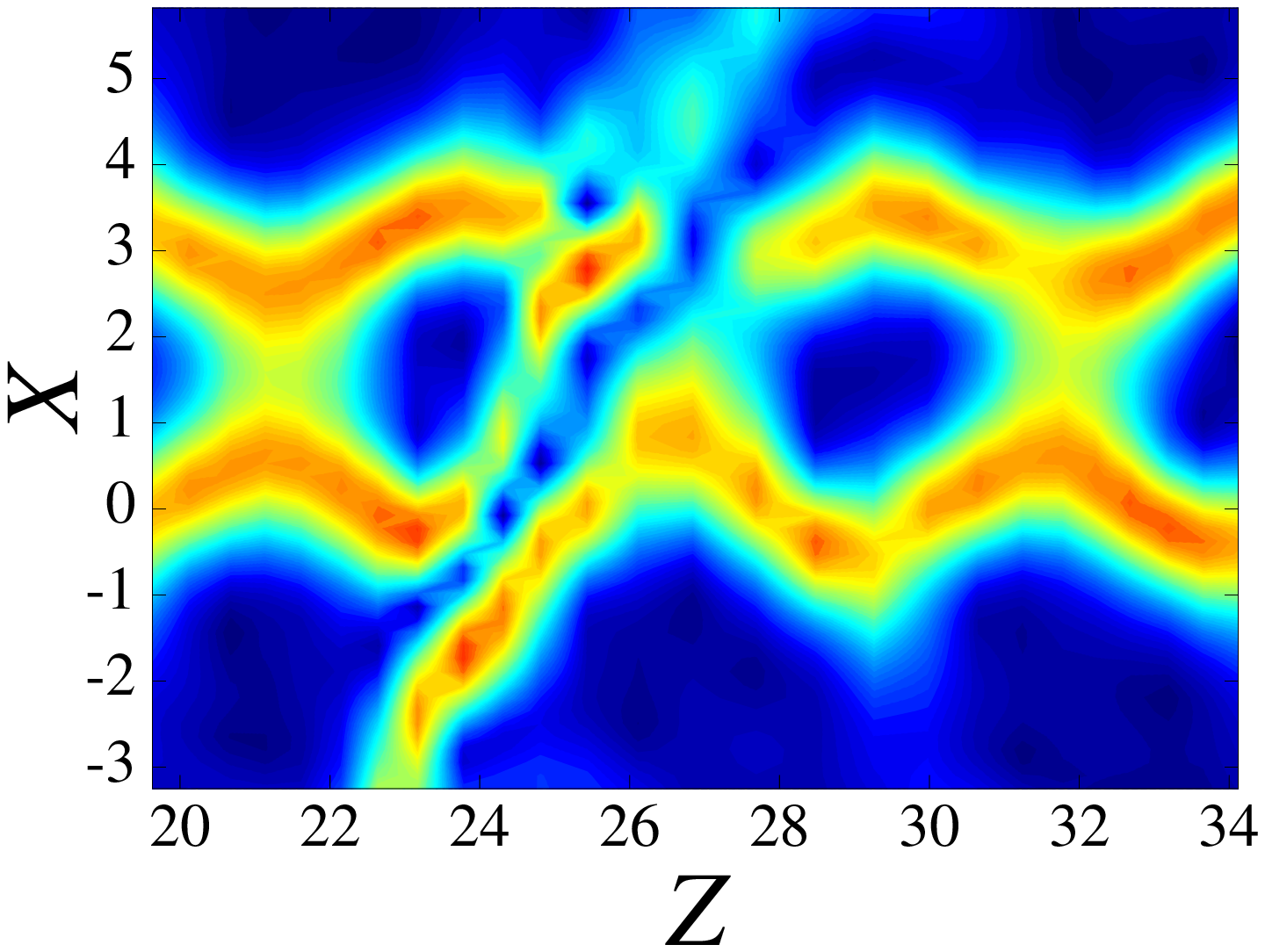}}
\end{center}
\caption{(Color online) (a) The cross section of field $\left\vert
u\left( X,Y,Z\right) \right\vert $ at $Y=0$, in the plane of $\left(
X,Z\right) $, for $k_{0}=1.816$. This is an example of the transient
regime, when the moving dipole is absorbed by the pair of trapped
ones after several quasi-elastic collisions. (b) The close-up of the
absorptive collision. The color code is the same as in Fig. 1a.}
\label{absdipolek01816theta0}
\end{figure}

\begin{figure}[th]
\begin{center}
\subfigure[]{\label{absdipolek01884theta0XvsZ}%
\includegraphics[width=5cm]{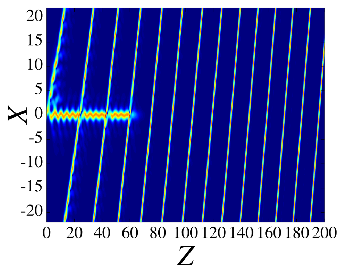}} %
\subfigure[]{\label{absdipolek01884theta0abs}%
\includegraphics[width=5cm]{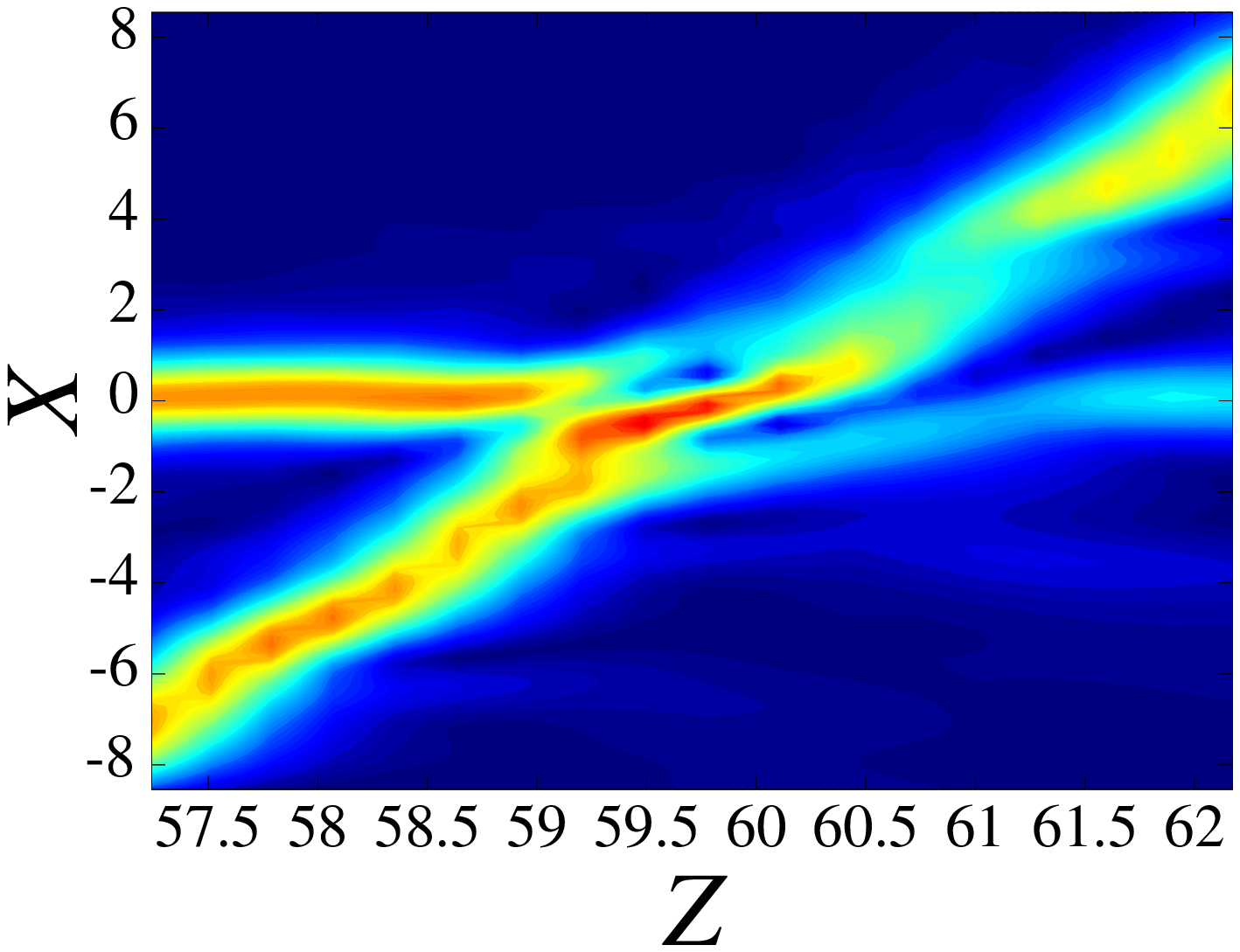}}
\end{center}
\caption{(Color online) (a) The cross section of field $\left\vert
u\left( X,Y,Z\right) \right\vert $ at $Y=0$, in the plane of $\left(
X,Z\right) $, for $k_{0}=1.884$. This is an example of
\textquotedblleft clearing the obstacle", when the moving dipole
absorbs the stationary one, after several collisions with it. (b)
The close-up of the absorptive collision. The color code is the same
as in Fig. 1a.} \label{absdipolek01884theta0}
\end{figure}
In other cases, the freely moving dipole is absorbed by the quiescent
pattern as a result of the first collision.

\begin{figure}[th]
\begin{center}
\subfigure[]{\label{absdipolek01884theta0sep}%
\includegraphics[width=5cm]{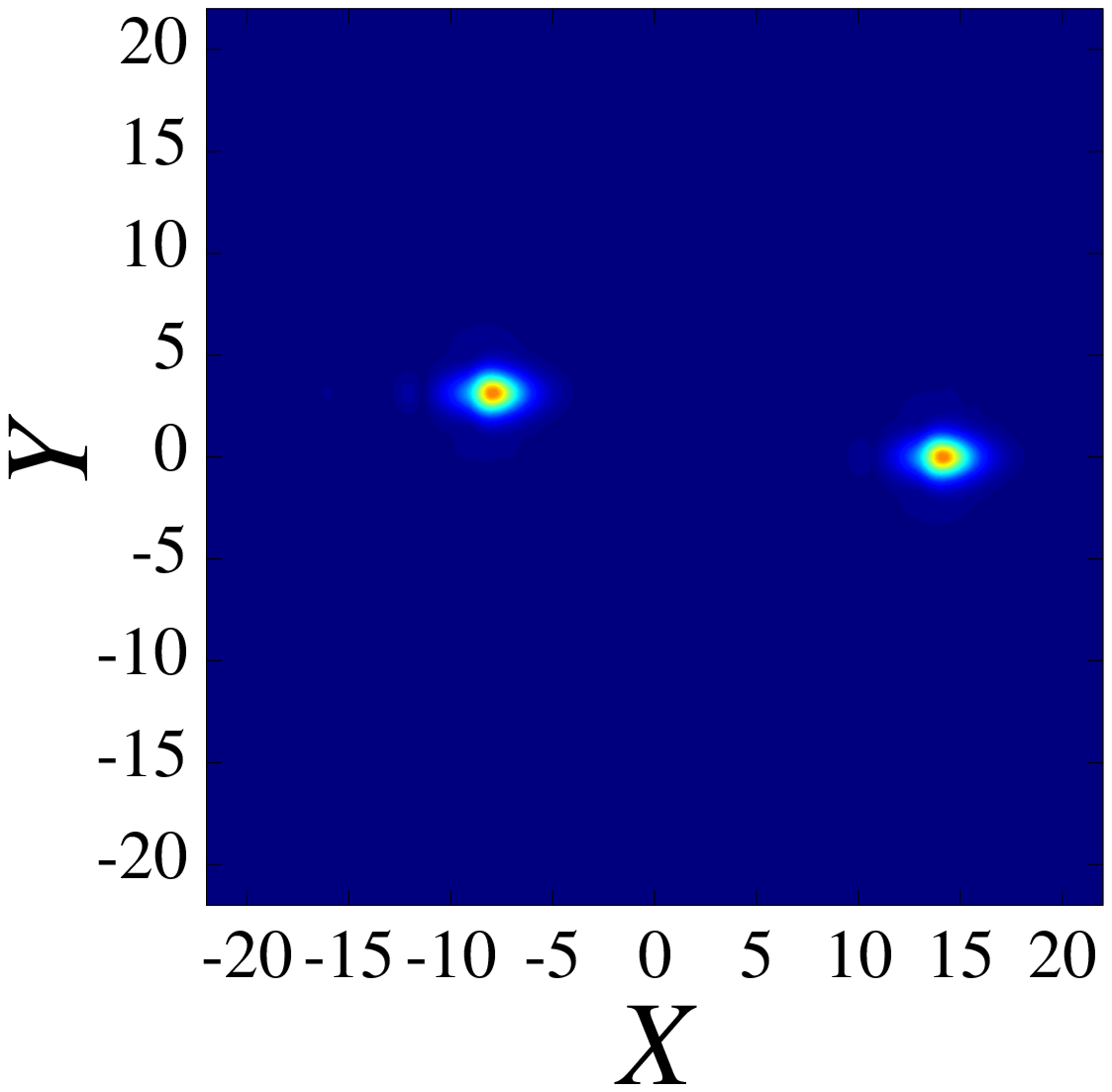}} %
\subfigure[]{\label{vitdipolek01884theta0}%
\includegraphics[width=5cm]{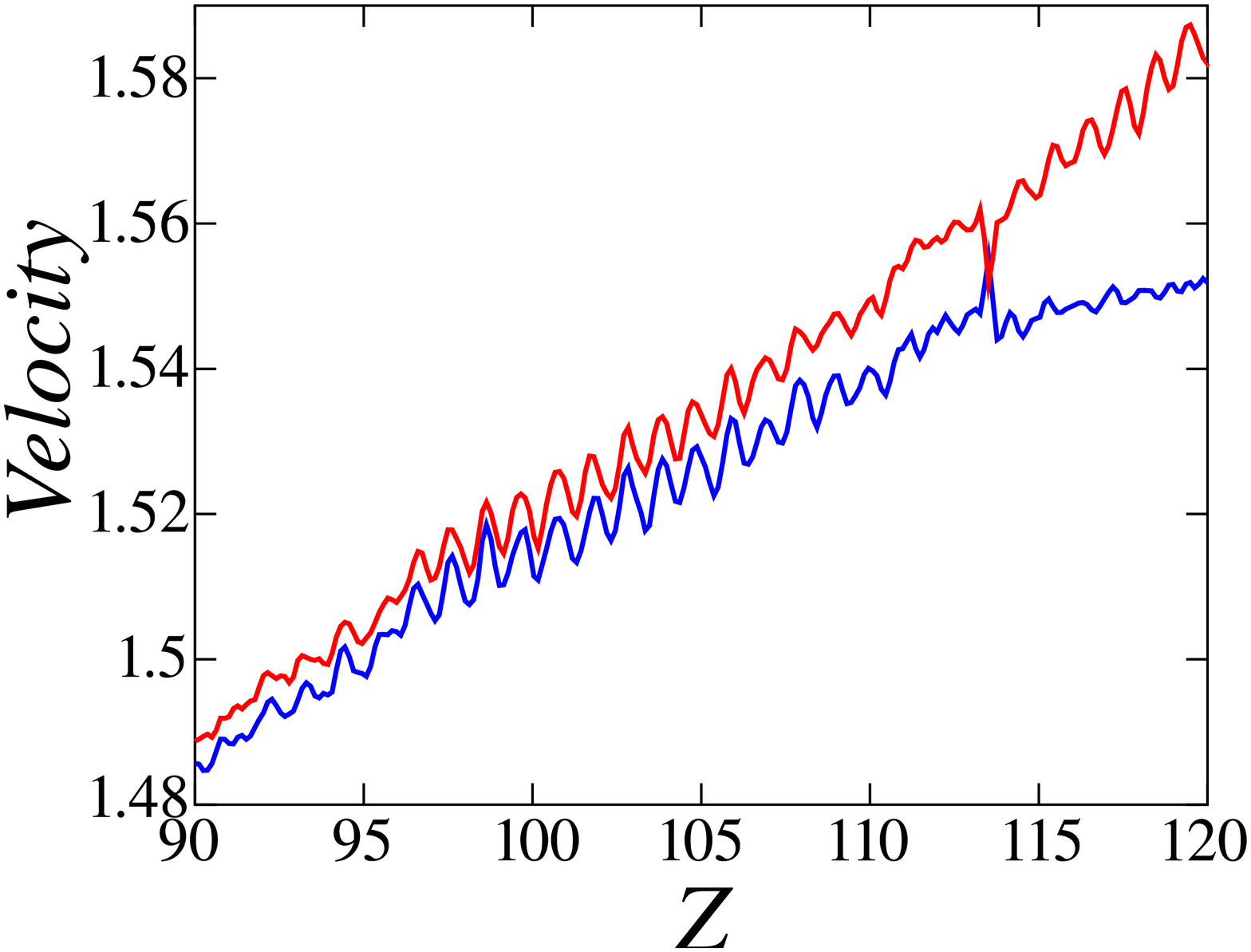}}\vfill
\subfigure[]{\label{absdipolek01884theta0sepy}%
\includegraphics[width=5cm]{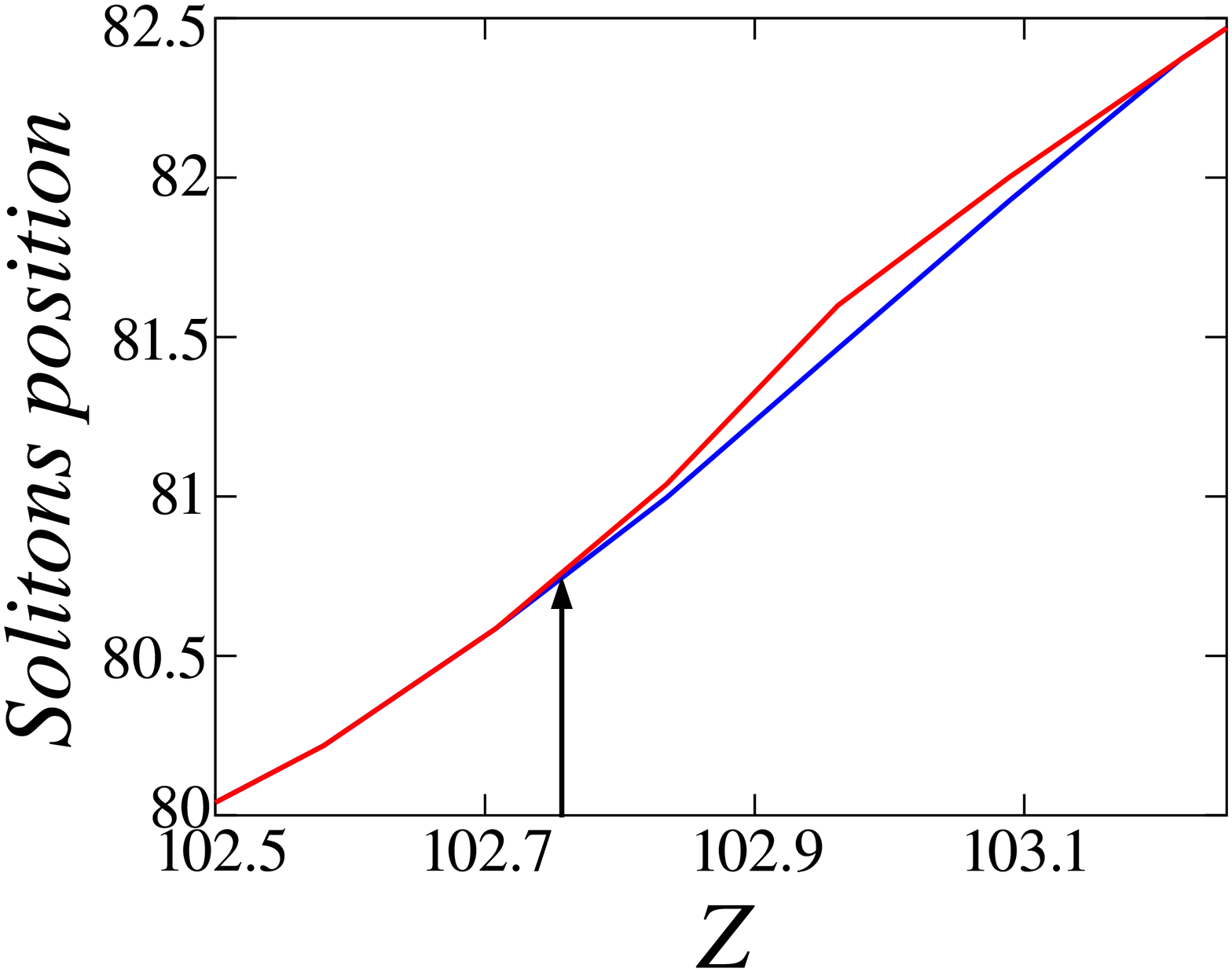}} %
\subfigure[]{\label{diffphasedipolek01884theta0}%
\includegraphics[width=5cm]{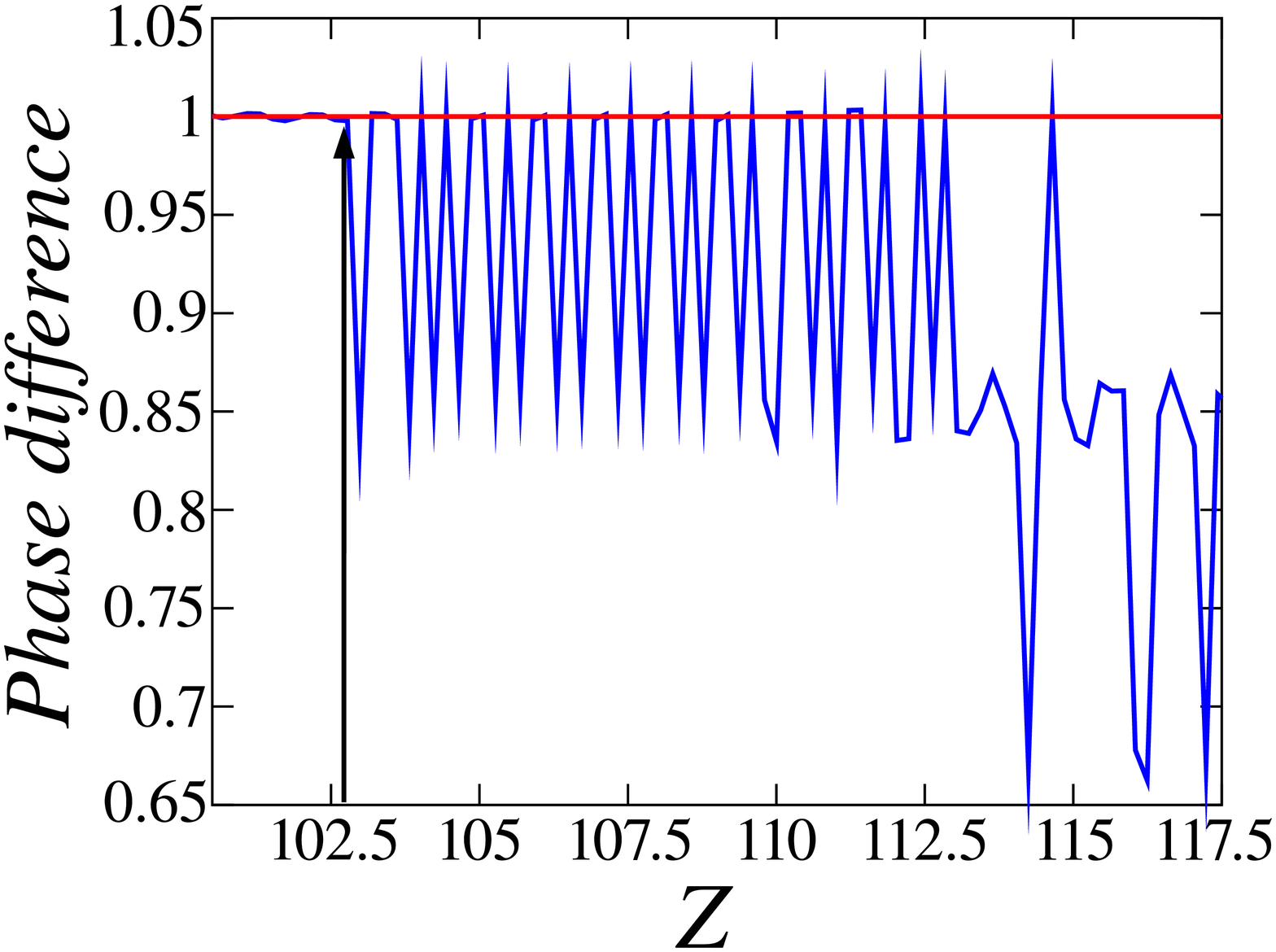}}
\end{center}
\caption{(Color online) Illustration of the splitting of the single
surviving dipole into uncorrelated fundamental solitons, which
follows \textquotedblleft clearing the obstacle", after the
absorption of the quiescent dipole by the moving one, at
$k_{0}=1.884$. (a) Field $\left\vert u\left( X,Y\right) \right\vert
$ at $Z=199.965$. The color code is the same as in Fig. 1a. (b,c).
Velocities and positions of both solitons as functions of $Z.$ (d)
The phase difference between the solitons versus $Z$, in units of
$\protect\pi $, the red
horizontal line corresponding to the phase difference equal to $\protect\pi $%
. The arrows in (c) and (d) indicate onset of the process which
eventually leads to the loss of the phase coherence and separation
of the two solitons.} \label{dipolek01884theta0abs}
\end{figure}

It is relevant to stress that, while the first two above-mentioned regimes
have been reported in Ref. \cite{cgl_mov} for the motion of kicked
fundamental solitons, the third regime (\textquotedblleft clearing the
obstacle") is a new one, which was not found for the fundamental solitons.
Another characteristic feature of the latter regime is that it eventually
leads to the splitting of the surviving single dipole into unbound
fundamental solitons, as shown in Fig. \ref{absdipolek01884theta0sep}.
To analyze the splitting, we have identified position $\left\{
X_{c},Y_{c}\right\} $ of the field maximum in each soliton (its center), and
values of phases at these points (mod $2\pi $), as functions of evolution
variable $Z$. As a result, it has been found that the splitting of the
dipole and the loss of the phase correlation between the splinters starts in
a \textquotedblleft latent form" at $Z\approx 102.8$, and becomes explicit
at $Z\simeq 112.5$, see Figs. \ref{absdipolek01884theta0sepy} and \ref%
{diffphasedipolek01884theta0}. The two splinter solitons get completely
separated at $Z\simeq 115$. The splitting also leads to the appearance of
the velocity difference between the solitons (the velocity is defined as $%
dX_{c}/dZ$), as seen in Fig. \ref{dipolek01884theta0abs}(b).

\section{The pattern formation by kicked quadrupoles}

A quadrupole is composed of four soliton-like power peaks, which are
mutually locked with phase difference $\pi $ between adjacent ones, see an
example of the offsite-centered (alias ``square-shaped") quadrupole in Fig. \ref%
{quadrinit}. Although this mode carries no vorticity, simulations
demonstrate that it is a very robust one. We here aim to investigate
dynamical regimes initiated by the application of the horizontal kick (\ref%
{kick}) to the quadrupole.

\begin{figure}[th]
\centering \subfigure[
]{\label{absquadrinit}\includegraphics[width=5cm]{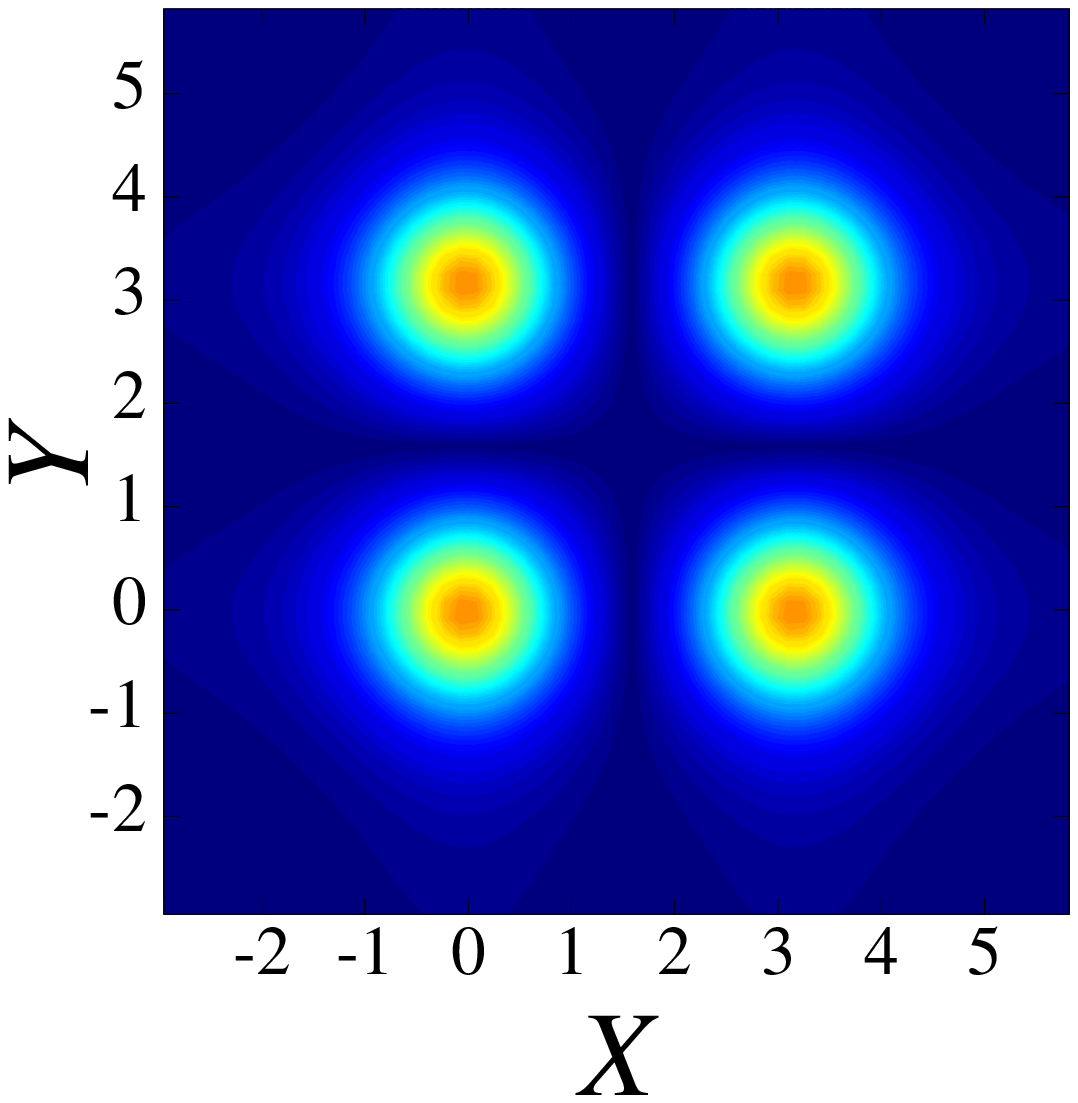}}
\subfigure[
]{\label{phasequadrinit}\includegraphics[width=5cm]{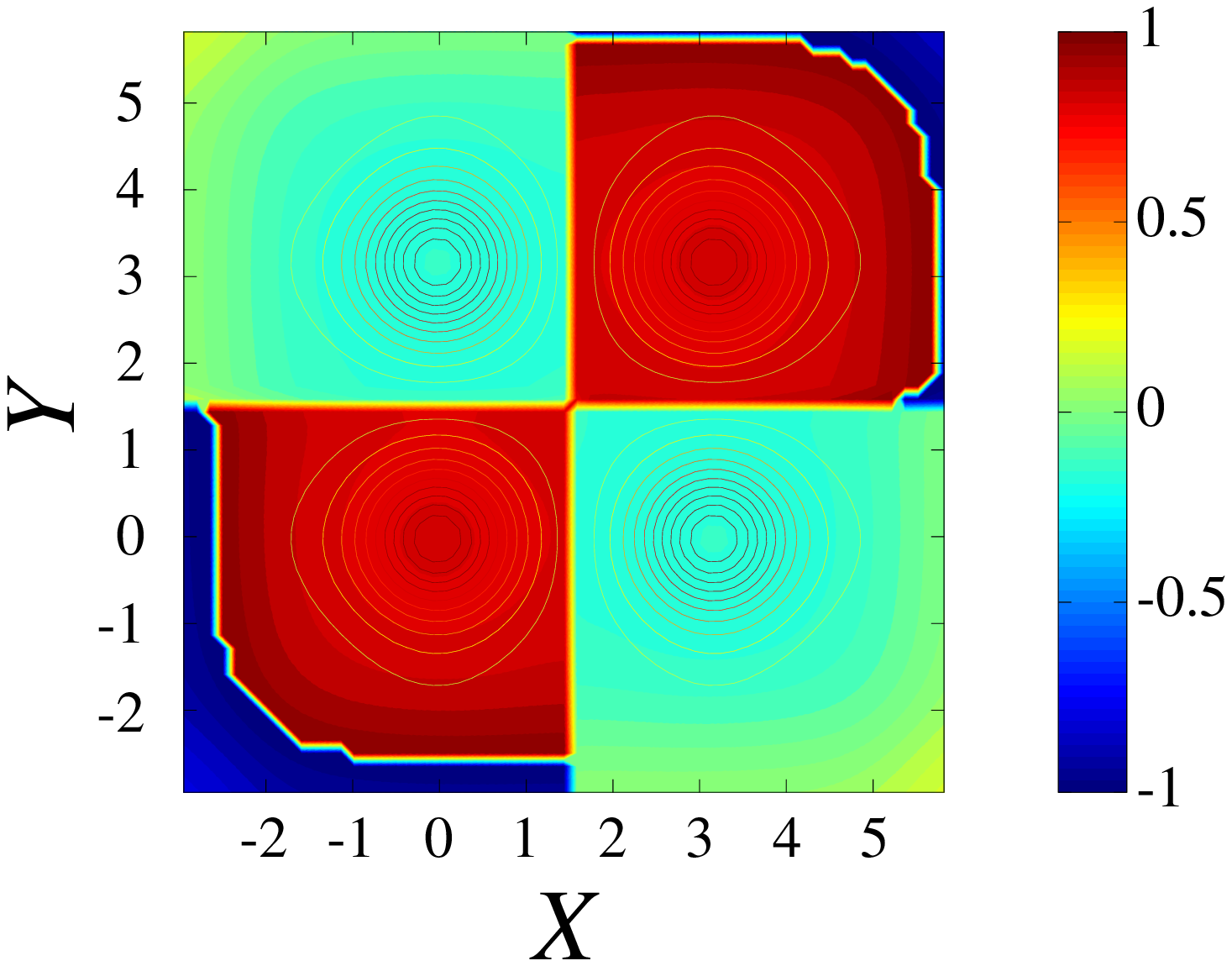}}
\caption{(Color online) The distribution of the amplitude (a) and
phase (in unites of $\protect\pi $) in the stable stationary
offsite-centered quadrupole used in the simulations. The color code
for the amplitude is the same as in Fig. 1a.} \label{quadrinit}
\end{figure}

\begin{figure}[th]
\begin{center}
\includegraphics[width=10cm]{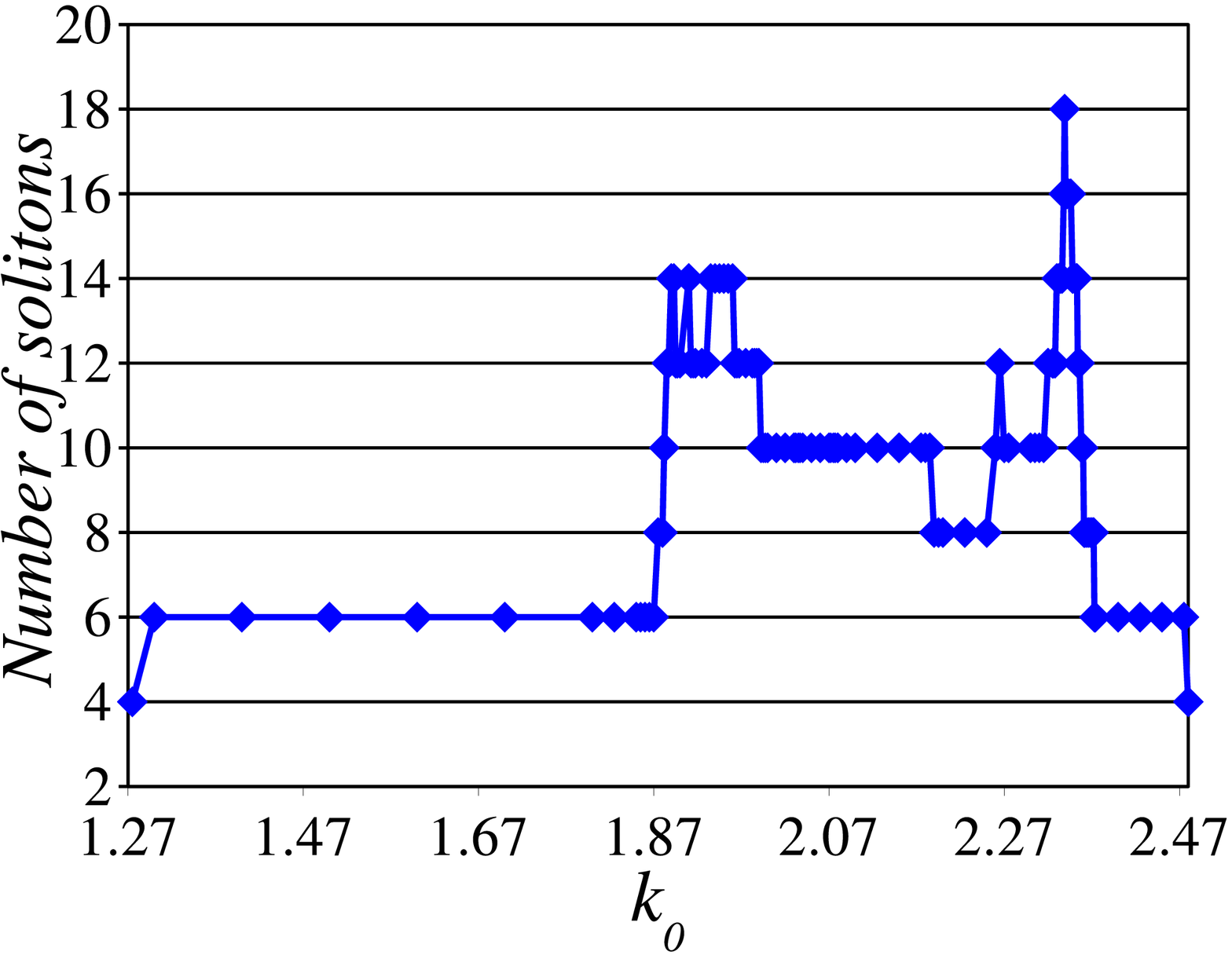}
\end{center}
\caption{(Color online) The total number of fundamental solitons in
the pattern produced by kick $k_{0}$ applied to the stable
offsite-centered quadrupole. Each dipole counts as two solitons.}
\label{evoquad}
\end{figure}

The quadrupole is set in motion by the kick whose strength exceeds the
respective threshold,
\begin{equation}
k_{0}^{(\mathrm{thr})}(\mathrm{quadr})=1.28,  \label{quadr}
\end{equation}%
cf. Eqs. (\ref{thr0}) and (\ref{thr_pi/2}). The horizontal motion of the
kicked quadrupole splits it into two vertical dipoles, and generates a set
of additional vertically arranged quiescent \textit{soliton pairs}, with a
phase shift of $\pi /2$ between them. The dependence of the total number of
solitons in the eventually established pattern on the kick's strength, $%
k_{0} $, is shown in Fig. \ref{evoquad}.
Because these simulations were subject to the periodic b.c.,
the free dipole completes the round trip to collide with the quiescent
pattern. The number of solitons was counted just before this collision.
In the case where there is no motion in the system (no free dipole emerges),
the count of the number of solitons is straightforward.

The result is quite different from that reported in the previous
section for the pattern formation by the kicked dipole, cf. Fig.
\ref{nombdipole}. Above the threshold value (\ref{quadr}), the
number of fundamental solitons in the emergent pattern increases and
remains constant in a wide interval of values of $k_{0}$,
\textit{viz}., six solitons for $k_{0}\in \lbrack 1.28,1.87]$. Then,
the number of the solitons increases to its maximum, which is 14 at
$k_{0}\in \lbrack 1.89,1.893]$, $k_0 = 1.91$ and $k_{0}\in \lbrack
1.935,1.96]$. Note that the increase is not monotonous. For example,
12 solitons are generated at $k_{0}=\lbrack 1.885,1.887]$ and
$k_{0}=\lbrack 1.895,1.9]$. Subsequently, in the interval of
$k_{0}\in \lbrack 1.9125,2.338] $, the soliton number varies between
8 and 16. The largest number of solitons, 18, is reached at
$k_{0}=2.339$. Then, the soliton number drops to 6, and this value
remains constant over a relatively broad interval, $k_{0}\in \lbrack
2.373,2.475]$. At still larger values of $k_{0}$, no additional
solitons are generated by the initially moving quadrupole, which in
this case again splits into two dipoles.

\begin{figure}[tbp]
\centering
\subfigure[]{\label{absquadrk02339theta0crea}%
\includegraphics[width=5cm]{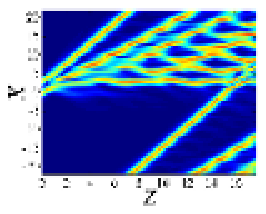}} %
\subfigure[]{\label{energquadrk02339theta0}%
\includegraphics[width=5cm]{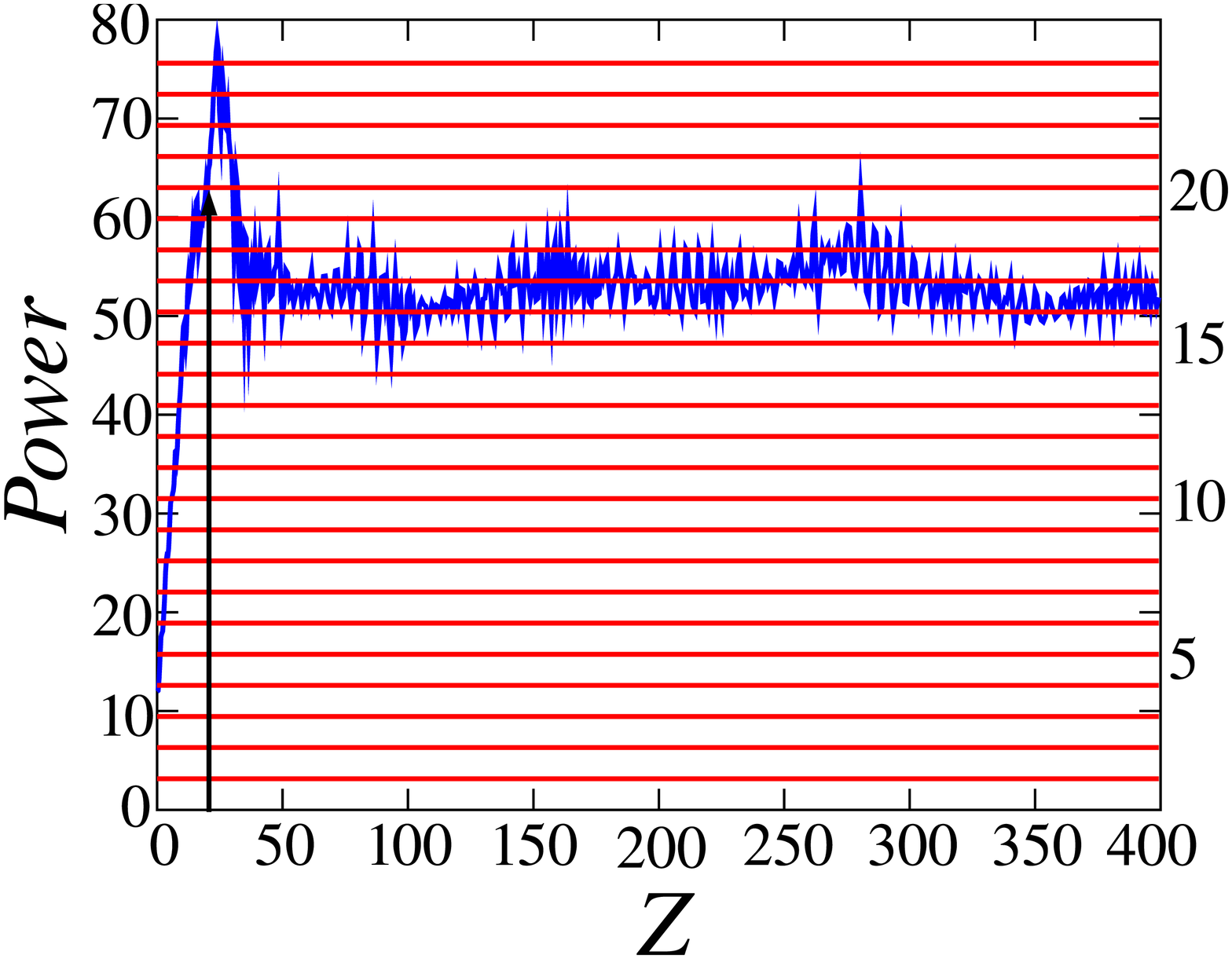}}\vfill
\subfigure[]{\label{absquadrk02339theta0max}%
\includegraphics[width=5cm]{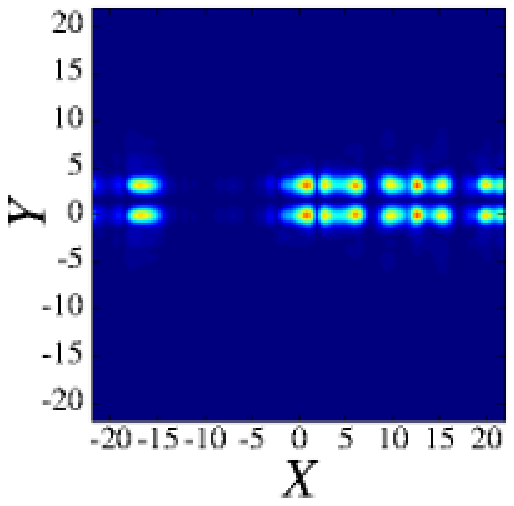}} %
\subfigure[]{\label{absquadrk02339theta0onde}%
\includegraphics[width=5cm]{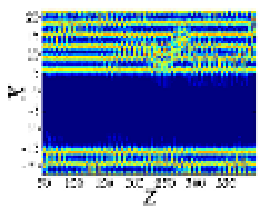}}
\caption{(Color online) The evolution of the horizontally kicked
quadrupole, for $k_{0}=2.339$. (a) Field $\left\vert u\left(
X,Y,Z\right) \right\vert $ in the cross section $Y=0$, before the
collision \ of the moving dipole with the pinned complex. (b) The
total power versus $Z$ (the vertical arrow indicates the collision
point); the horizontal red lines show the power corresponding to $n$
quiescent fundamental solitons, $n$ being the numbers indicated on
the right vertical axis.(c) Field $\left\vert u\left( X,Y\right)
\right\vert $ at $Z=399.34$. (d) Field $\left\vert u\left(
X,Y,Z\right) \right\vert $ in the cross section $Y=0$, after the
collision. The color code is the same as in Fig. 1a.}
\label{quadrk02339theta0}
\end{figure}

At $k_{0}=2.339$, the simulations generate a set of 18 solitons (the
largest number, as said above). At first, two moving dipoles
are actually produced by the splitting of the original quadrupole, see Fig. %
\ref{absquadrk02339theta0crea}. The faster dipole [whose trajectory
is characterized by a larger slope (velocity), $dX_{c}/dZ$] moves
without creating new solitons, while the slower one creates several
of them, namely, the third moving dipole and six quiescent ones,
which brings the total number of solitons to 18, as said above. The
total energy increases up to about 24 times the energy of a
quiescent soliton, which corresponds to the 12 such solitons, plus
the 3 moving dipoles, with the energy of a moving soliton being
about twice that of a quiescent one (see Fig.
\ref{energquadrk02339theta0}). Due to the periodic b.c., the three
moving dipoles hit the previously generated quiescent chain, one
after the other (see Fig. \ref{absquadrk02339theta0crea}). As a
result, two first dipoles are captured by the chain increasing the
number of the bound solitons in it, while the third moving dipole is
absorbed without adding new solitons to the chain. This complex
interaction results in a chain of 8 quiescent dipoles (equivalent to
16 solitons). The so generated dipole train originally features
intrinsic oscillations, which are eventually damped, see Fig.
\ref{absquadrk02339theta0onde}. Note that Fig.
\ref{absquadrk02339theta0crea} shows only the constituent
fundamental solitons on line $Y=0$, in terms of Fig. \ref%
{absquadrk02339theta0max}, their counterparts on the line of $Y=3$
showing the same picture.

\begin{figure}[tbp]
\centering
\includegraphics[width=10cm]{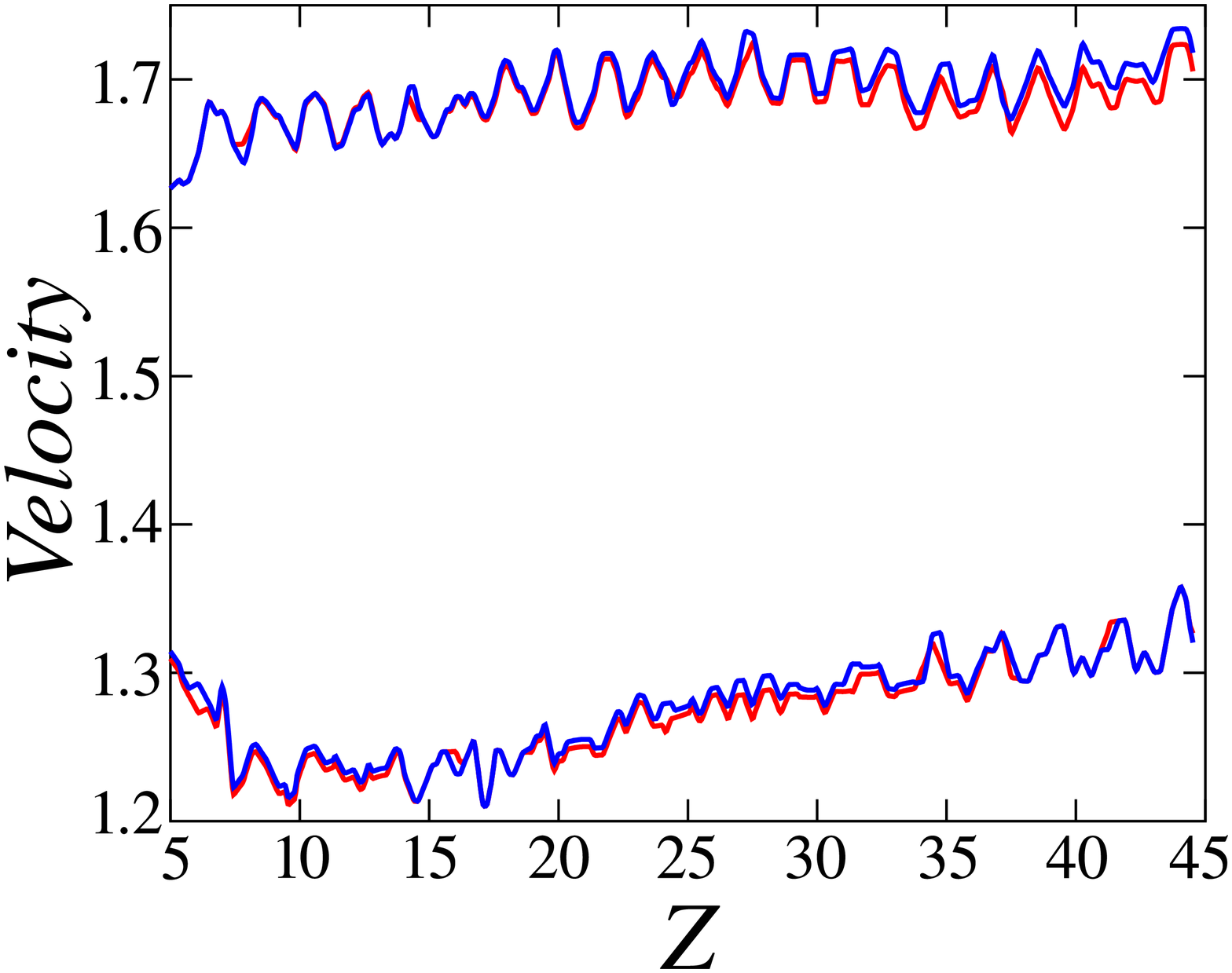}
\caption{(Color online) Velocities of two dipoles into which the kicked
quadrupole splits at $k_{0}=3$.}
\label{vitquadrk03theta0}
\end{figure}

As mentioned above and shown in Fig. \ref{vitquadrk03theta0}, at $k_{0}>2.48$
the initial quadrupole splits into two dipoles, which move at different
velocities, without the formation of additional soliton pairs.
Each dipole keeps the phase difference
of $\pi $ between the constituent solitons.

\section{The pattern formation by kicked vortices}

\subsection{Chaotic patterns generated by kicked rhombic (onsite-centered) vortices}

\begin{figure}[th]
\centering
\subfigure[]{\label{absrhombusinit}%
\includegraphics[width=5cm]{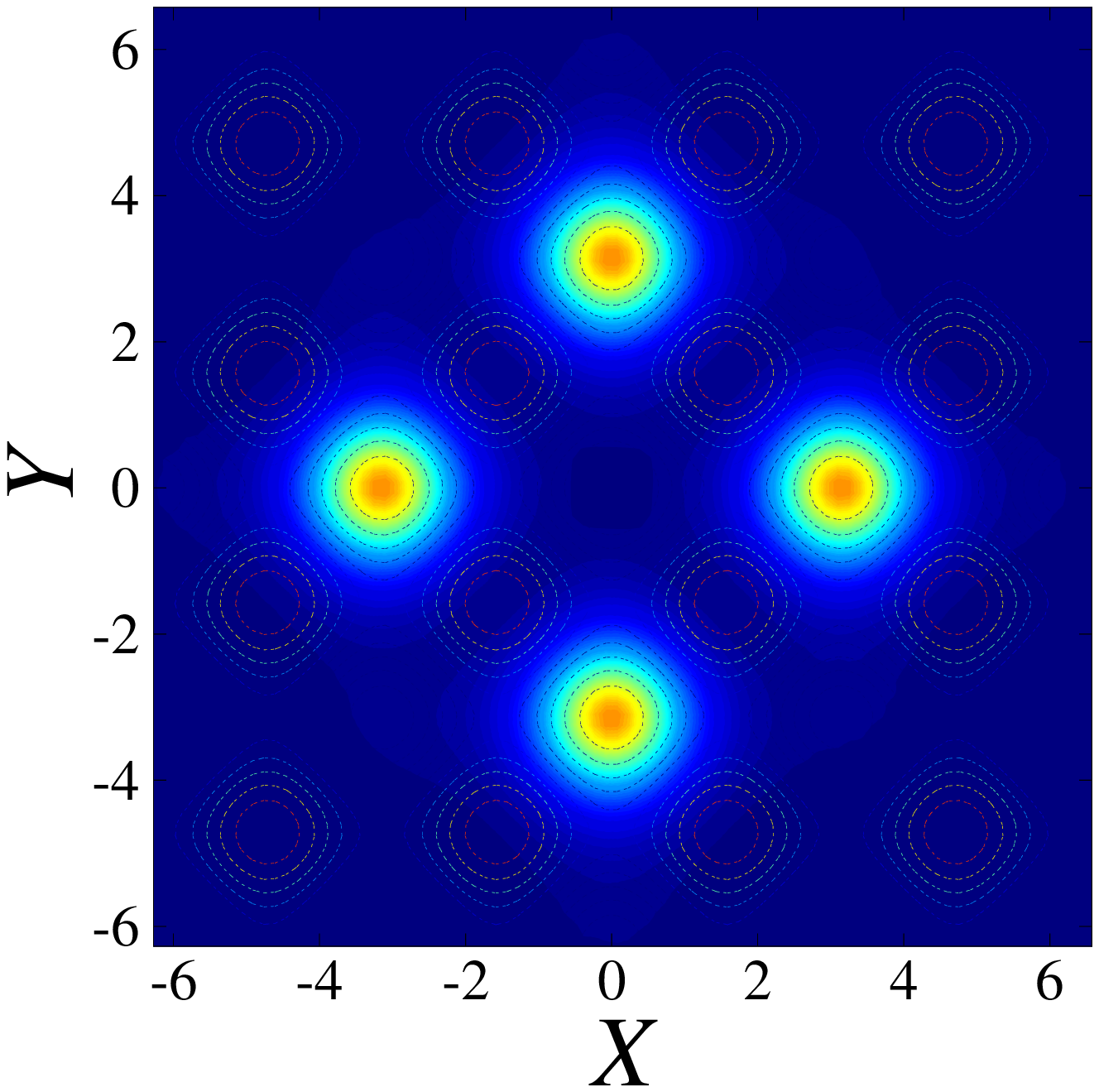}} \subfigure[]{%
\label{phaserhombusinit}\includegraphics[width=5cm]{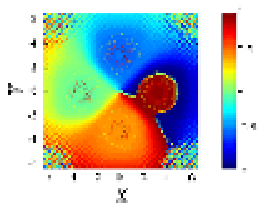}}
\caption{(Color online) (a) and (b): The distribution of the
amplitude and phase (in units of $\protect\pi $) in the stable
onsite-centered (rhombus-shaped) vortex. The lines are level
contours of potential $V$. The color code for the amplitude is the
same as in Fig. 1a.} \label{rhombiqueini}
\end{figure}

It is well known that the lattice potential supports localized
vortical modes of two types, the onsite- and offsite-centered ones
\cite{BBB,Thawatchai,Yang}. First, we consider the pattern-formation
dynamics for horizontally kicked rhombic vortices built of four
fundamental solitons with an empty site in the center, which carry
the
total phase circulation of $2\pi $, corresponding to the topological charge $%
S=1$, see Fig. \ref{absrhombusinit}.

\begin{figure}[th]
\centering
\subfigure[
]{\label{diffphaserhombusstab}\includegraphics[width=5cm]{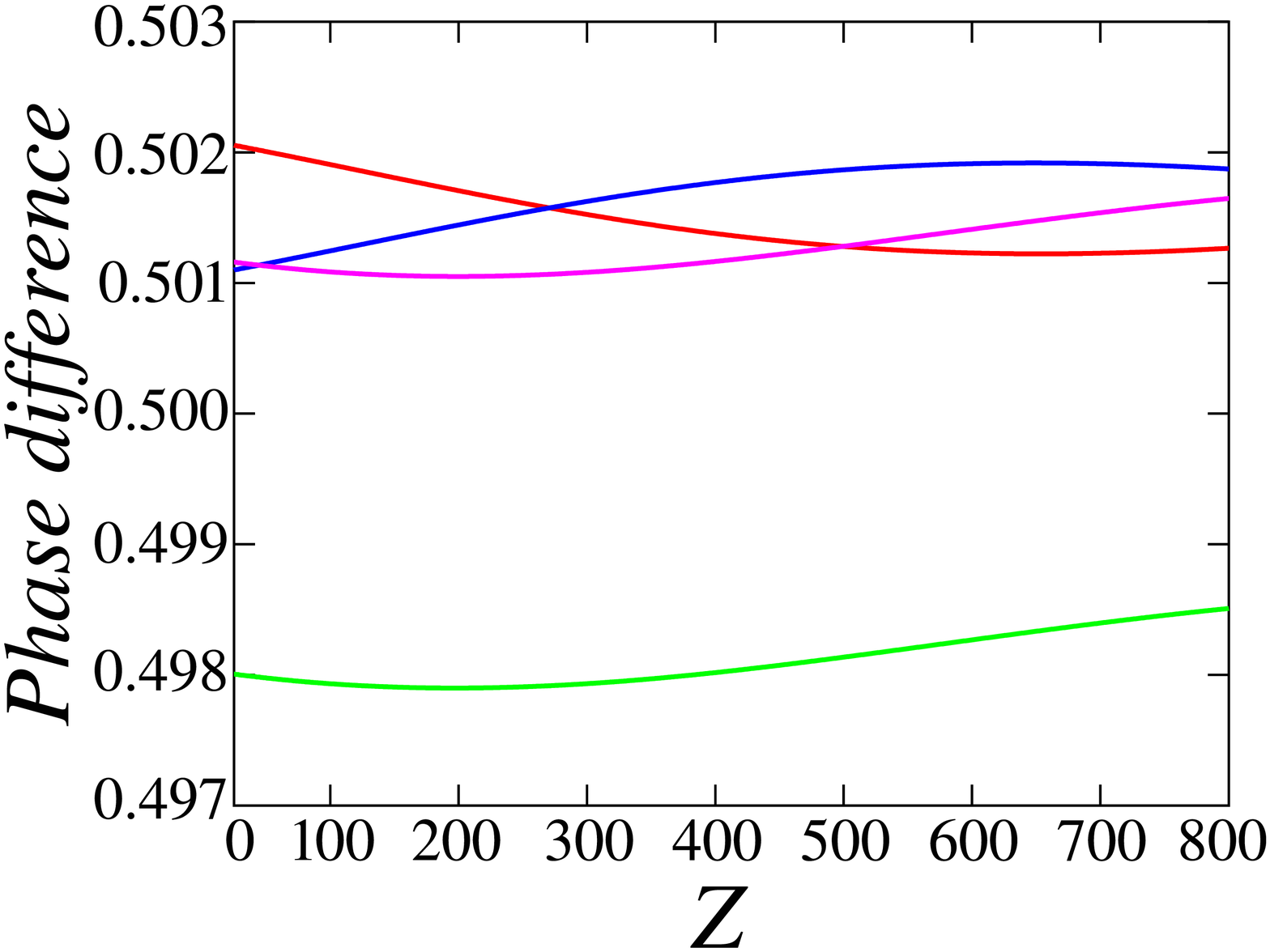}}
\subfigure[
]{\label{diffphaserhombusk001}\includegraphics[width=5cm]{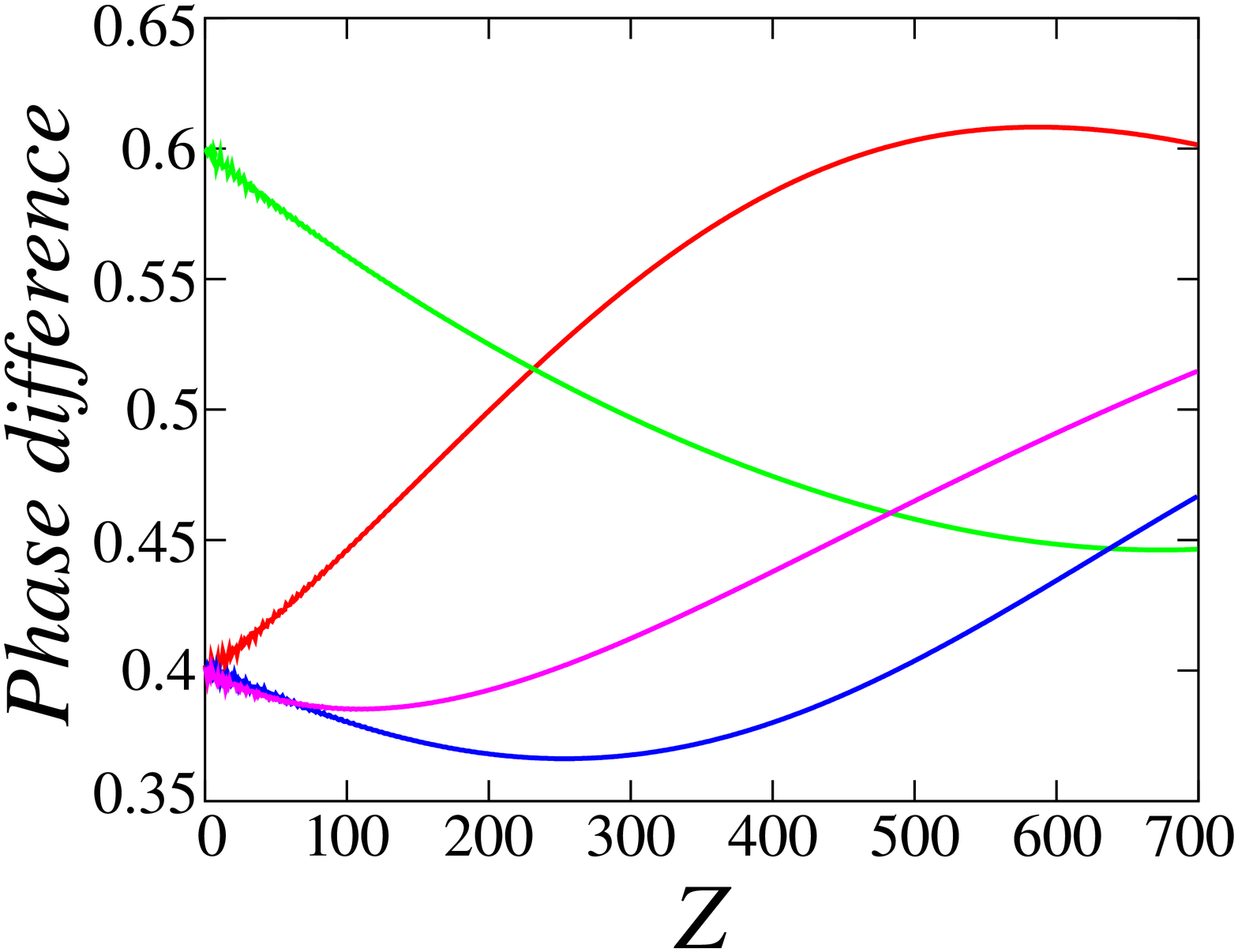}}%
\vfill
\subfigure[
]{\label{diffphaserhombusk002}\includegraphics[width=5cm]{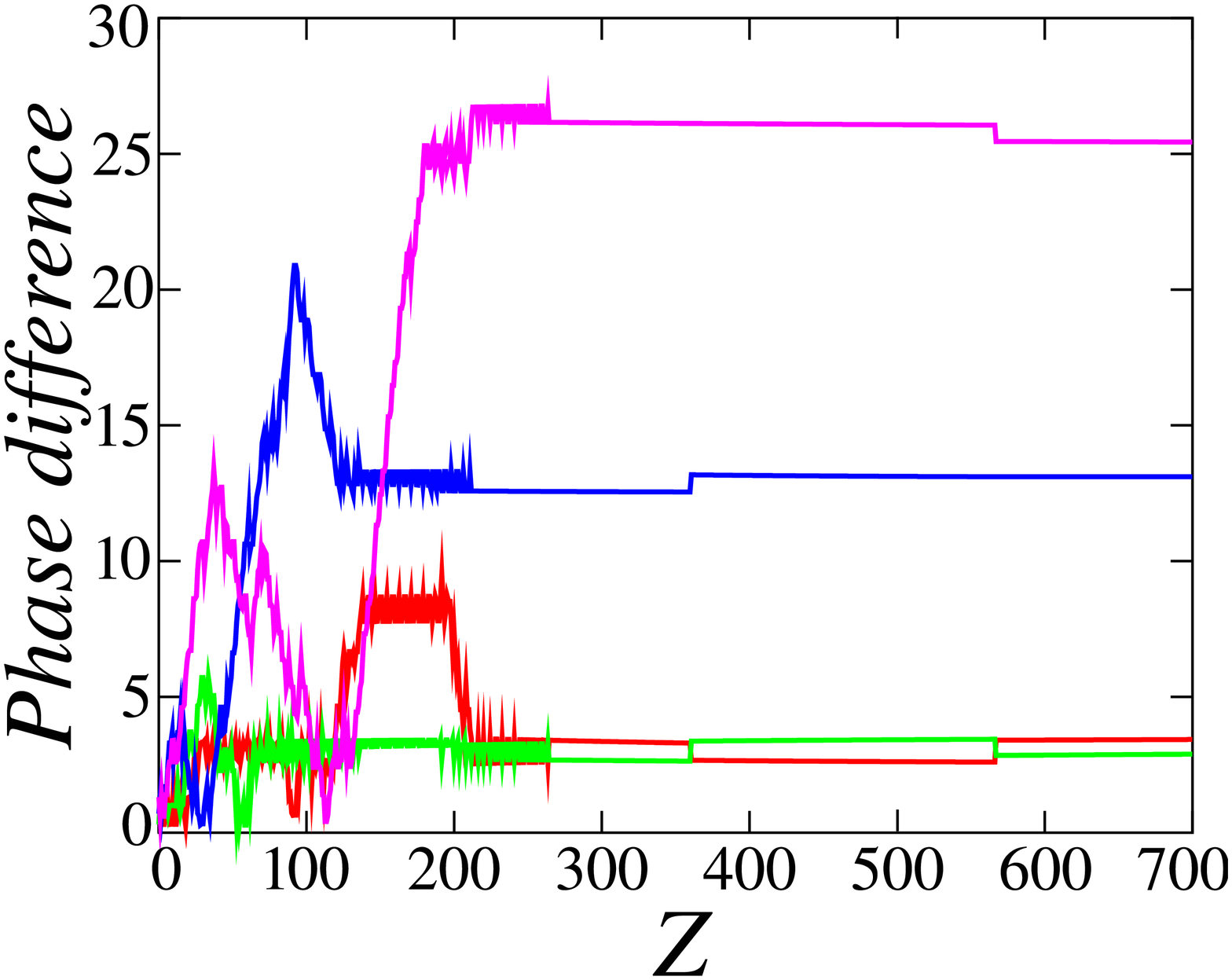}}
\caption{(Color online) The phase difference between adjacent constituent
solitons (in units of $\protect\pi $), versus $Z$, in a weakly kicked
rhombic vortex, {for different values of the kick's strength: (a) $k_{0}=0$,
(b) $k_{0}=0.1$, (c) $k_{0}=0.2$.} }
\label{diffphaserhombus}
\end{figure}

A weak horizontal kick, with $k_{0}\lesssim 0.1$, excites oscillations of
the constituent fundamental solitons which built the vortex, while vorticity
$S=1$ is kept (i.e., phase differences between the adjacent solitons remain
very close to $\pi /2$), see Fig. \ref{diffphaserhombusk001}. A stronger
kick (for instance, $k_{0}=0.5$) destroys the vortical phase structure, and
transforms the vortex into a quadrupole, as shown in Fig. \ref%
{diffphaserhombusk002}.

\begin{figure}[th]
\centering
\includegraphics[width=10cm]{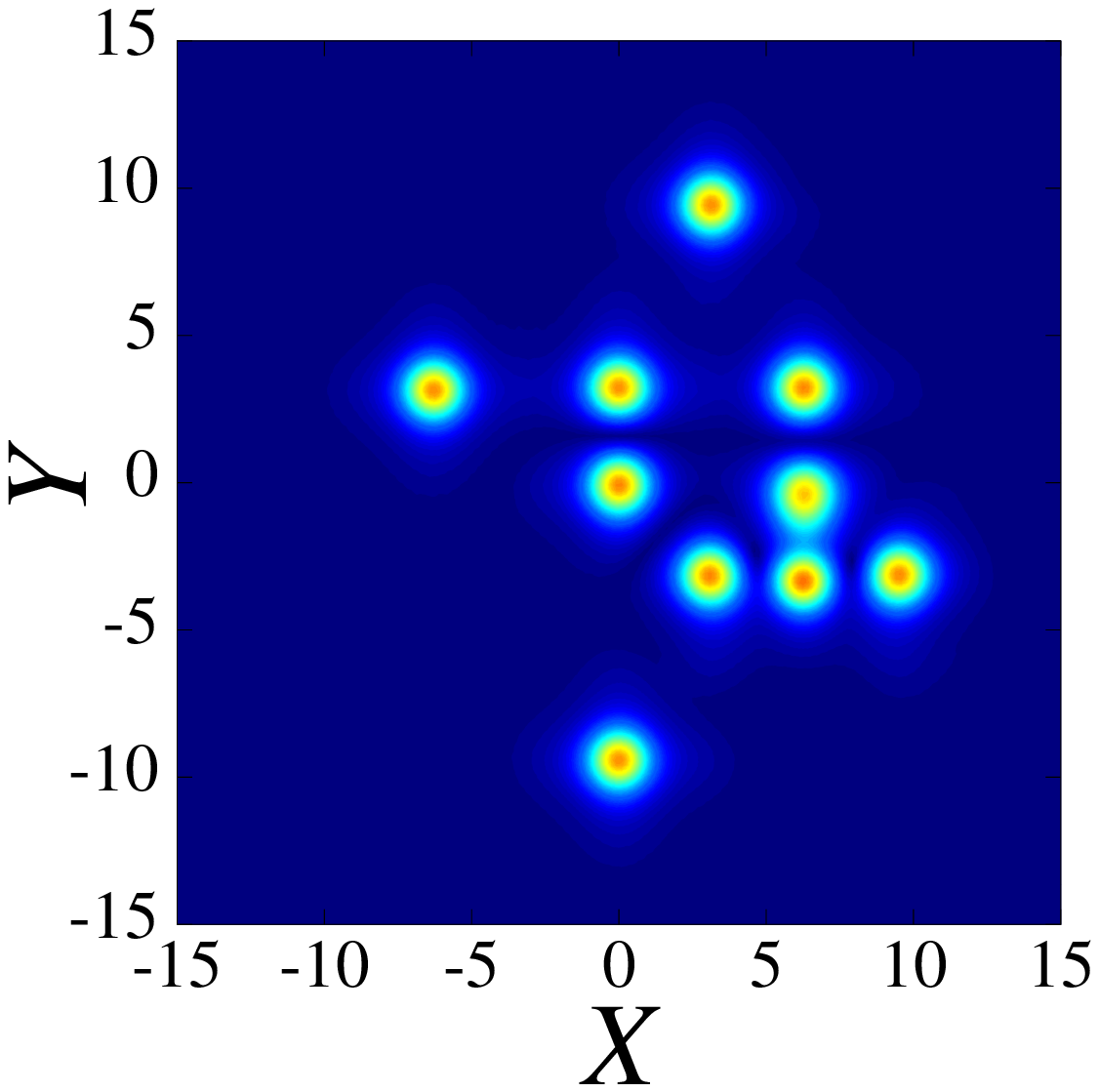}
\caption{(Color online) Field $\left\vert u\left( X,Y\right)
\right\vert $ at $Z=299.725$, generated by the kicked rhombic vortex
for $k_{0}=1.0$. The color code is the same as in Fig. 1a.}
\label{rho_k0_10_theta_amp}
\end{figure}

\begin{figure}[th]
\centering
\includegraphics[width=10cm]{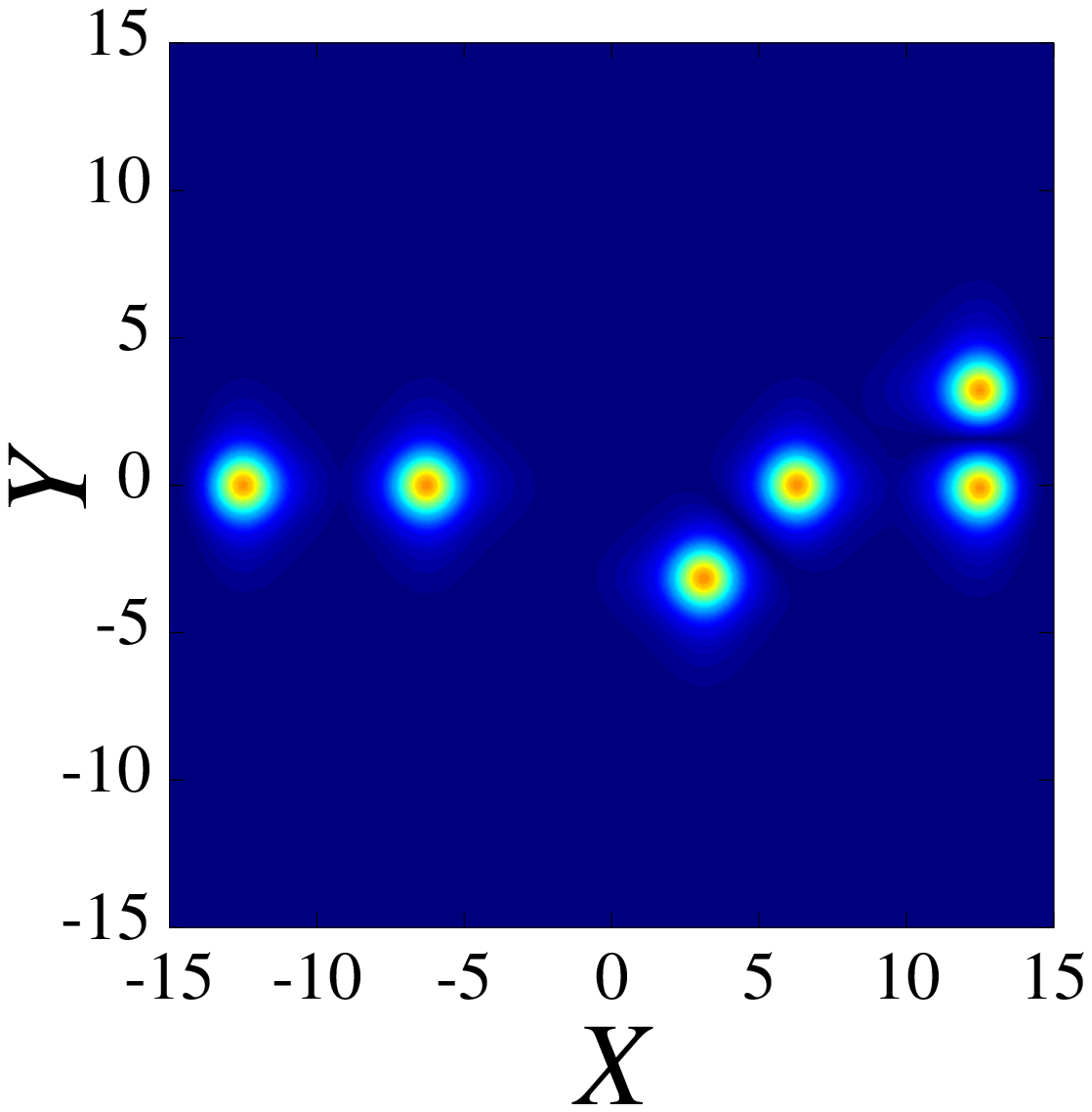}
\caption{(Color online) The same as in Fig. \protect\ref{rho_k0_10_theta_amp}%
, but for $k_{0}=1.5$. The color code is the same as in Fig. 1a.}
\label{rho_k0_15_theta_amp}
\end{figure}

At $k_{0}=1.0$ and $k_{0}=1.5$, see Figs. \ref{rho_k0_10_theta_amp} and \ref%
{rho_k0_15_theta_amp}, respectively, the kick completely destroys the
vortices, which are replaced by apparently random clusters of quiescent
fundamental solitons. Note that, although the results shown in Figs. \ref%
{rho_k0_10_theta_amp}-\ref{rho_k0_15_theta_amp} have been obtained with
absorbing b.c., rather than periodic ones, this circumstance does not affect
the results. The same type of b.c. is used below.

\subsection{Kicked offsite-centered (square-shaped) vortices}

\begin{figure}[th]
\centering
\subfigure[]{\label{abssquareinit}%
\includegraphics[width=5cm]{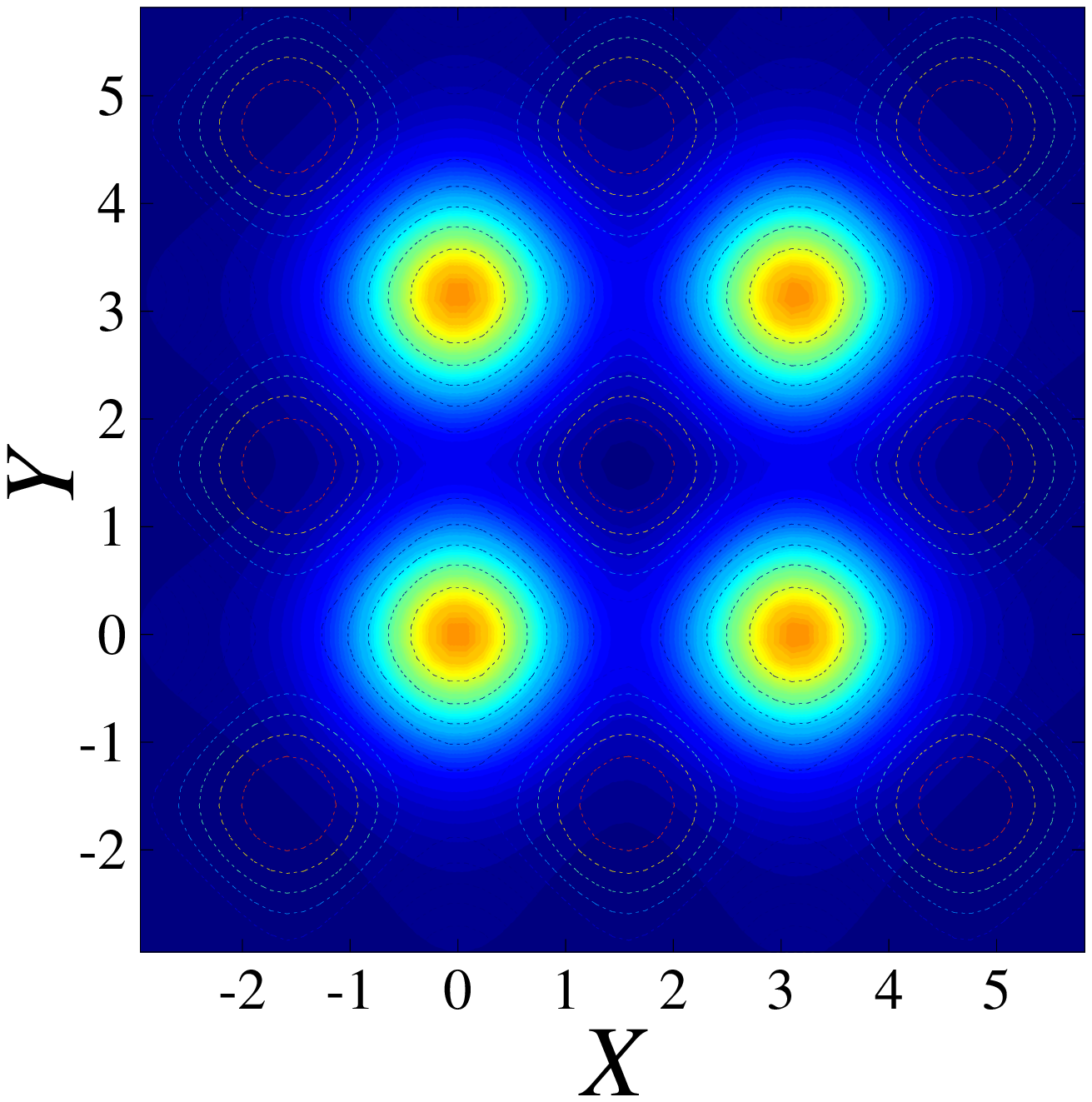}} \subfigure[]{%
\label{phasesquareinit}\includegraphics[width=5cm]{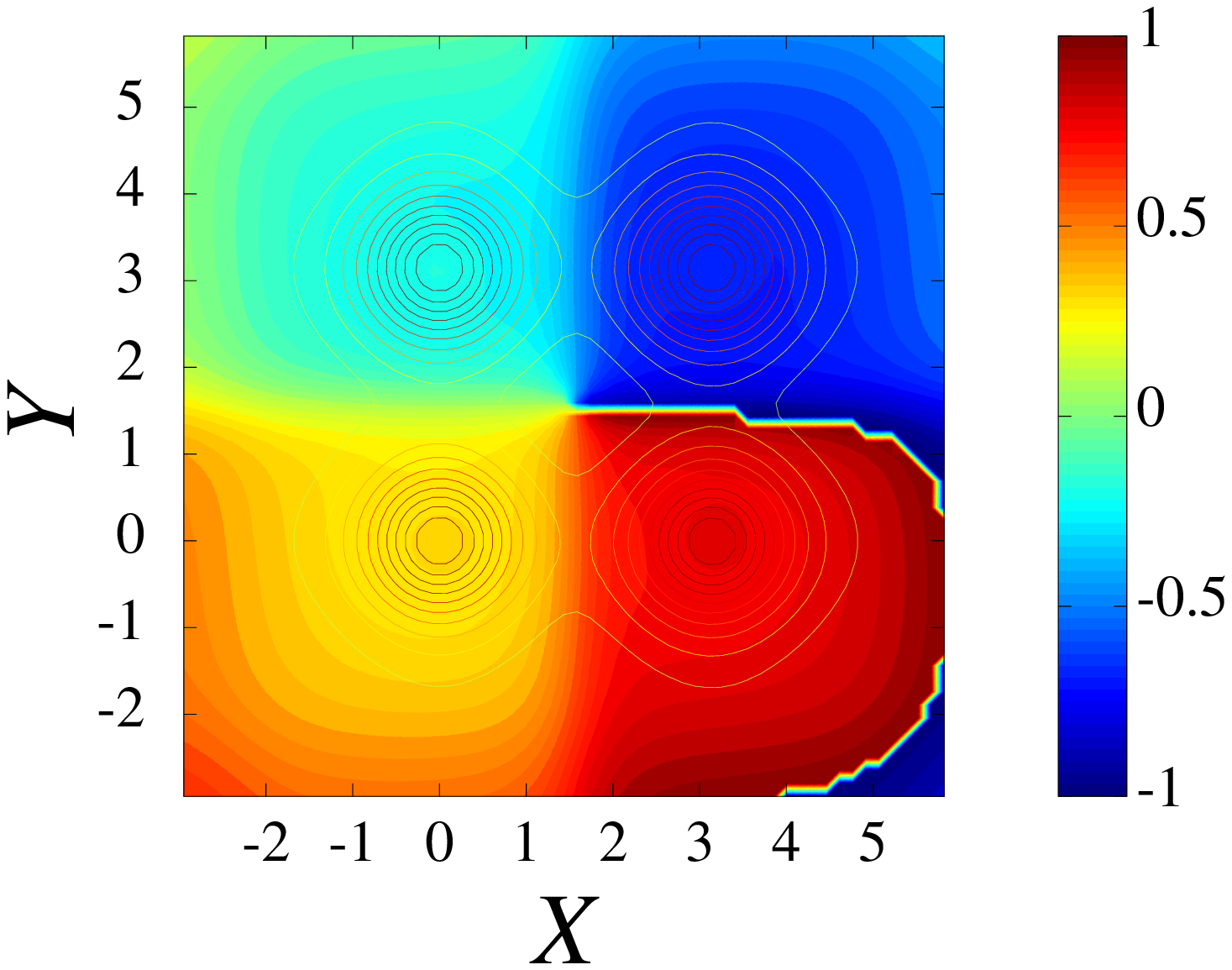}}
\caption{(Color online) The distribution of the amplitude (a) and
phase (b) in the unstable offsite-centered vortex. The lines are
level contours of potential $V$. The color code for the amplitude is
the same as in Fig. 1a.} \label{squareinit}
\end{figure}

Unlike their onsite-centered counterparts, quiescent
offsite-centered vortices, such as the one shown in Fig.
\ref{squareinit}, are unstable in the entire parameter space of Eq.
(\ref{CGL}) which we have explored, in agreement with the general
trend of the offsite-centered vortices to be more fragile than their
onsite-centered counterparts \cite{Thawatchai}. As a result of the
instability development, they are transformed into stable
quadrupoles. Nevertheless, results displayed below confirm that it
is relevant to consider dynamical pattern formation by unstable
kicked vortices as this type.

First, we consider the application of the horizontal kick (\ref{kick})
corresponding to $\theta =0$ and varying strength $k_{0}$. The fundamental
solitons building the vortex oscillate without setting in progressive motion
below the threshold, $k_{0}\leq k_{0}^{\mathrm{(thr)}}=1.2125$, cf. Eqs. (%
\ref{thr0}), (\ref{thr_pi/2}), and (\ref{quadr}). Actually, it may
happen, in this case, that a new soliton is created and starts
moving in the horizontal direction, but the energy is not sufficient
to stabilize it, and the new soliton decays eventually, while the
initial solitons which compose the offsite-centered vortex are
recovered at the original positions. The inner phase structure of
the unstable offsite-centered vortices is destroyed in the course of
the oscillations, and it transforms into a quadrupole, in accordance
with the above-mentioned fact that this is the outcome of its
instability in the absence of the kick.

\begin{figure}[th]
\centering
\subfigure[$Z=2.6593$]{%
\includegraphics[width=5cm]{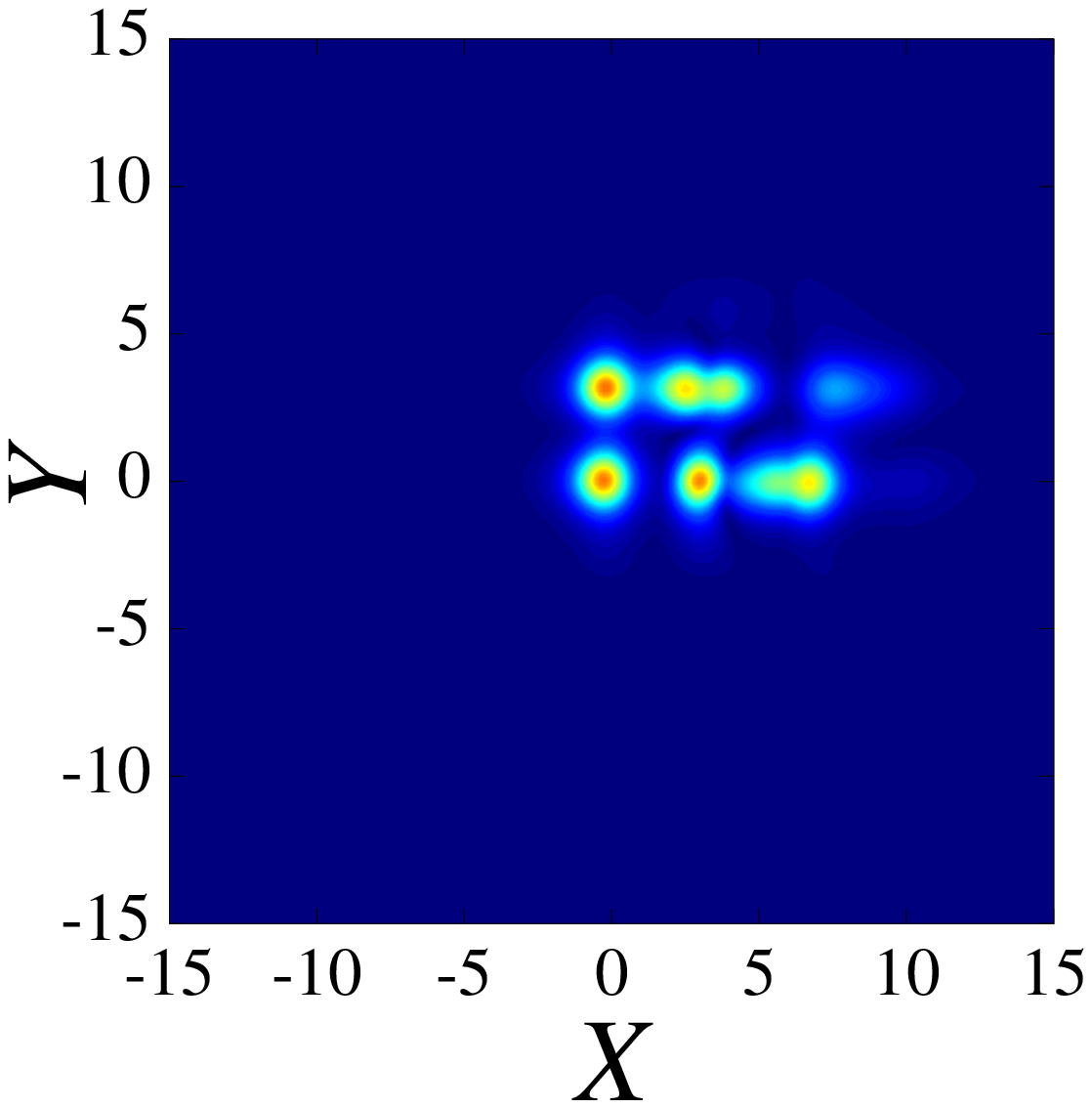}}%
\subfigure[$Z=12.008$]{%
\includegraphics[width=5cm]{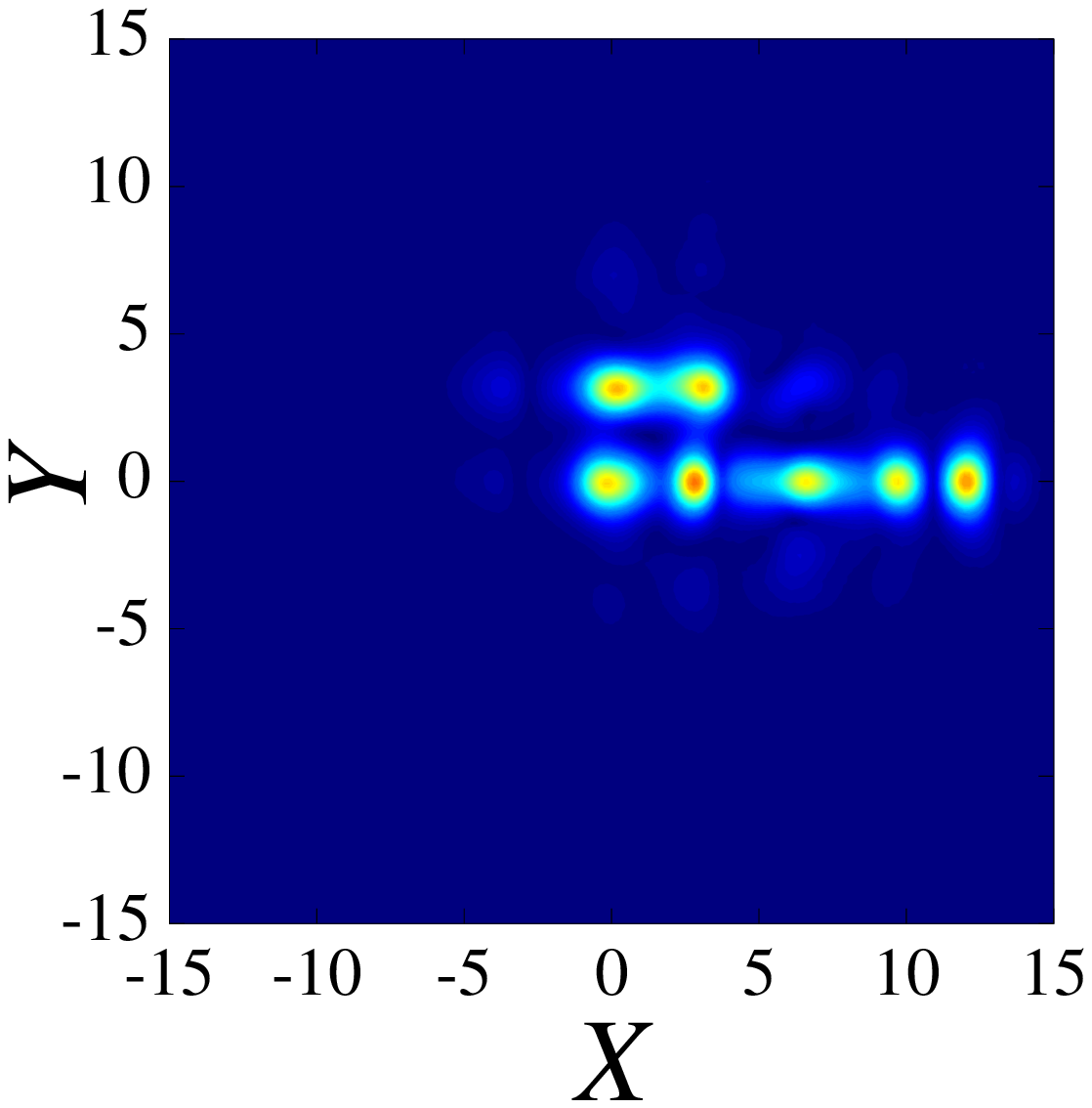}}\vfill%
\subfigure[$Z=239.91$]{%
\includegraphics[width=5cm]{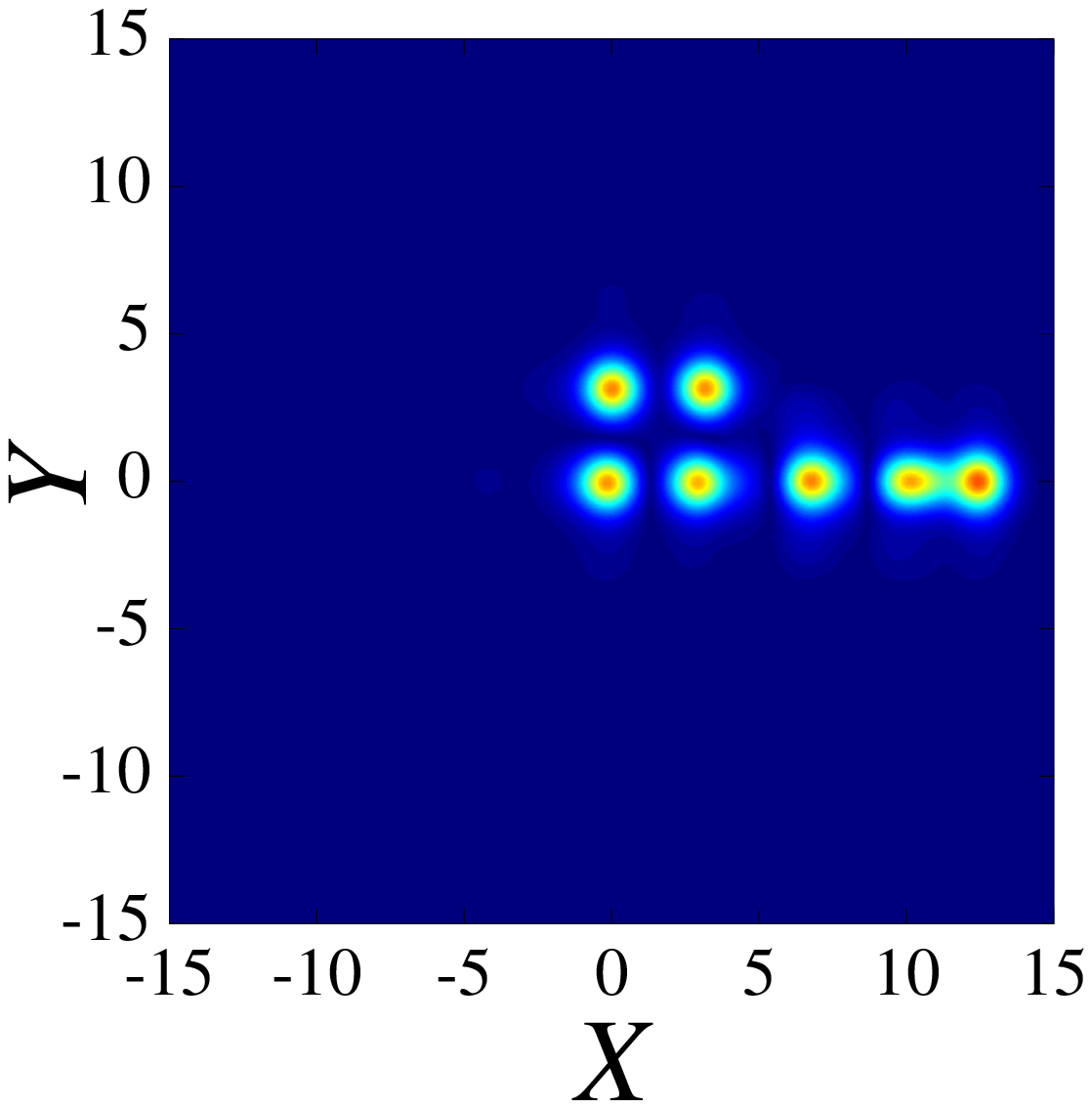}} %
\subfigure[$Z=299.825$]{%
\includegraphics[width=5cm]{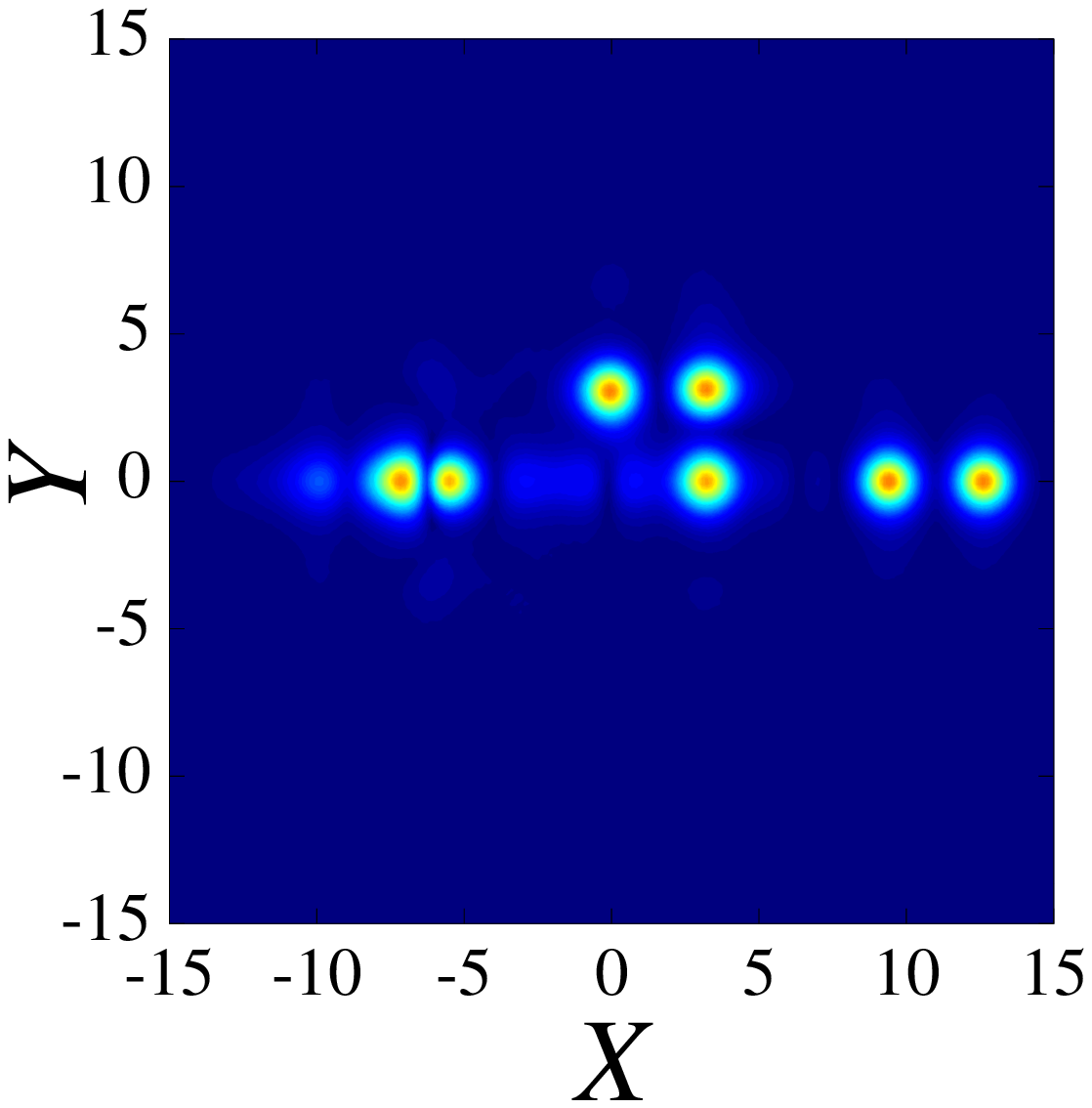}}
\caption{(Color online) The evolution of the unstable offsite-centered
vortex kicked in the horizontal direction ($\protect%
\theta =0$) with $k_{0}=1.5$. The color code is the same as in Fig.
1a.} \label{sqc_k0_15}
\end{figure}

\begin{figure}[th]
\centering
\includegraphics[width=10cm]{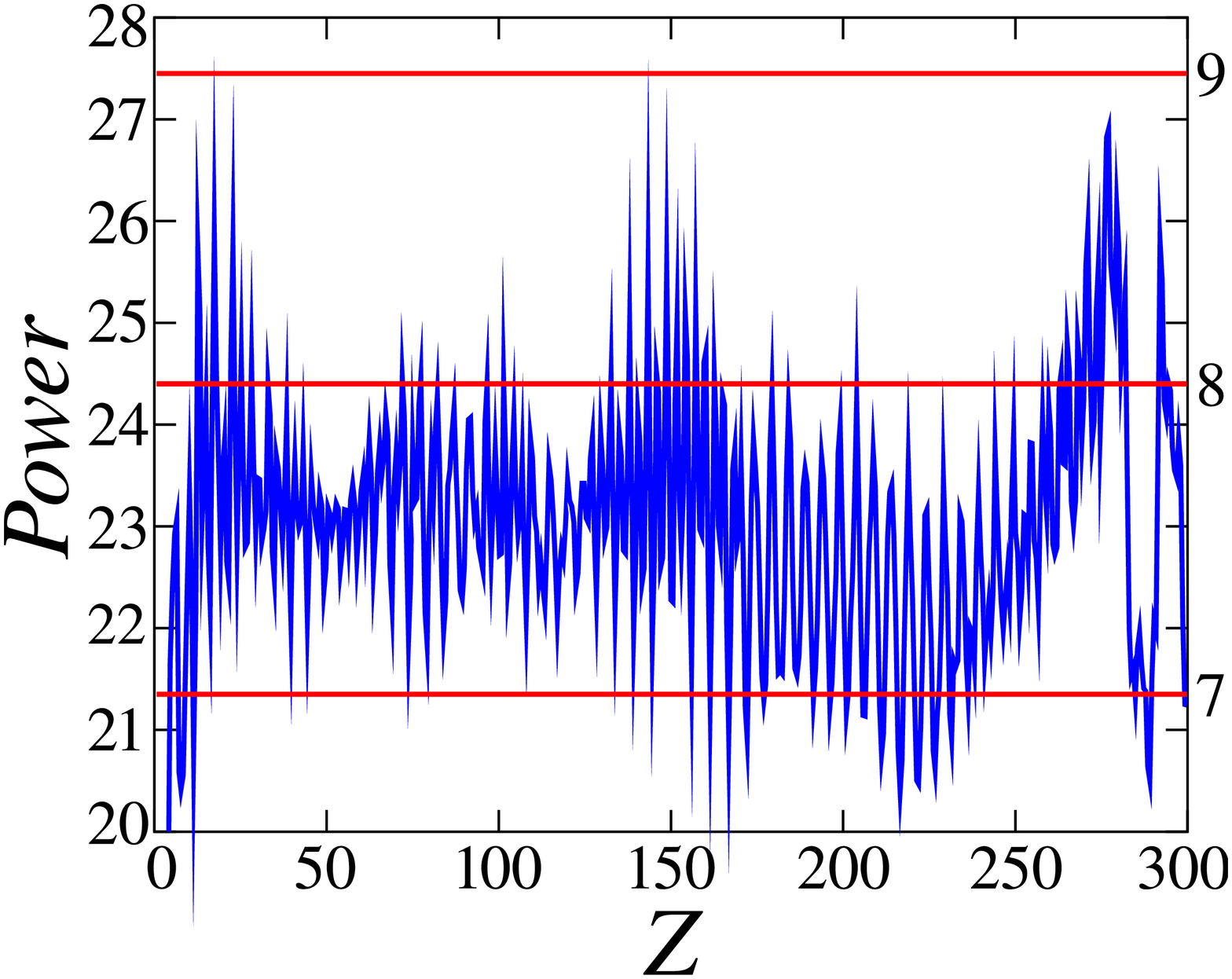}
\caption{(Color online) The evolution of the total power for the
pattern produced by horizontally kicking the offsite-centered
vortex, for $k_{0}=1.5$. The red horizontal lines show power levels
corresponding to $n$ quiescent solitons.} \label{sqc_k0_15_energ}
\end{figure}

The increase of $k_{0}$ leads to formation of new 2D patterns. At $k_{0}=1.5$%
, the right vertical pair (column) of the fundamental solitons, which are a
part of the original vortical square, start to duplicate themselves, while
moving to the right (in the direction of the kick), see Fig. \ref{sqc_k0_15}%
. A noteworthy effect is breaking of the symmetry between the top and bottom
solitons in the column by the kick, only the bottom soliton succeeding to
create a horizontal array of additional solitons (three ones, in total).%
\textbf{\ }In this case, Fig. \ref{sqc_k0_15_energ} shows that the eventual
value of the total power (\ref{P}) oscillates between values corresponding
to the cumulative power of $7$ or $8$ quiescent fundamental solitons. The
resulting pattern develops a disordered form, which keeps to oscillate
randomly, as Fig. \ref{sqc_k0_15_energ} clearly demonstrate.

\begin{figure}[th]
\centering
\subfigure[$Z=1.0784$]{%
\includegraphics[width=5cm]{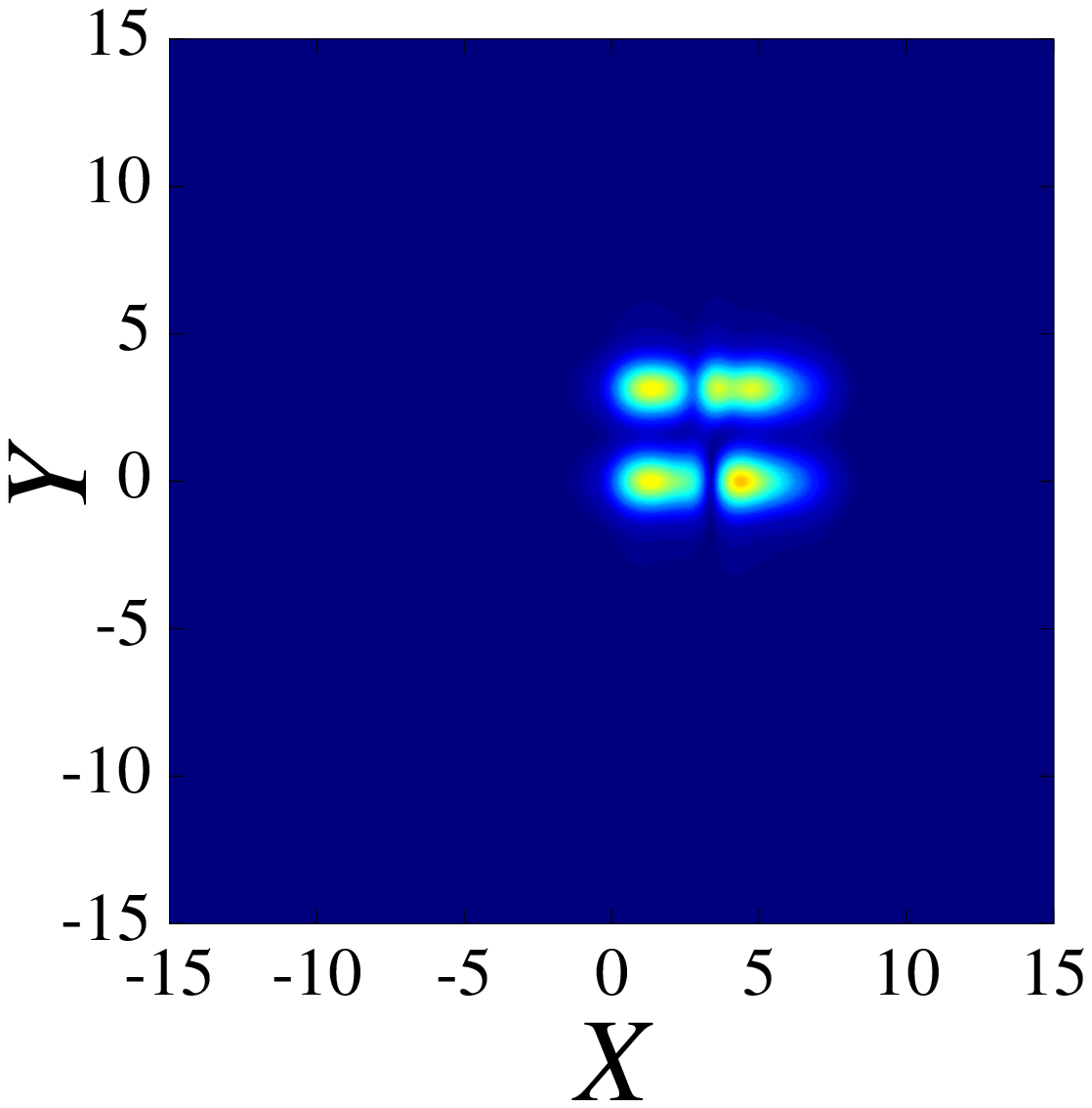}} %
\subfigure[$Z=2.0680$]{%
\includegraphics[width=5cm]{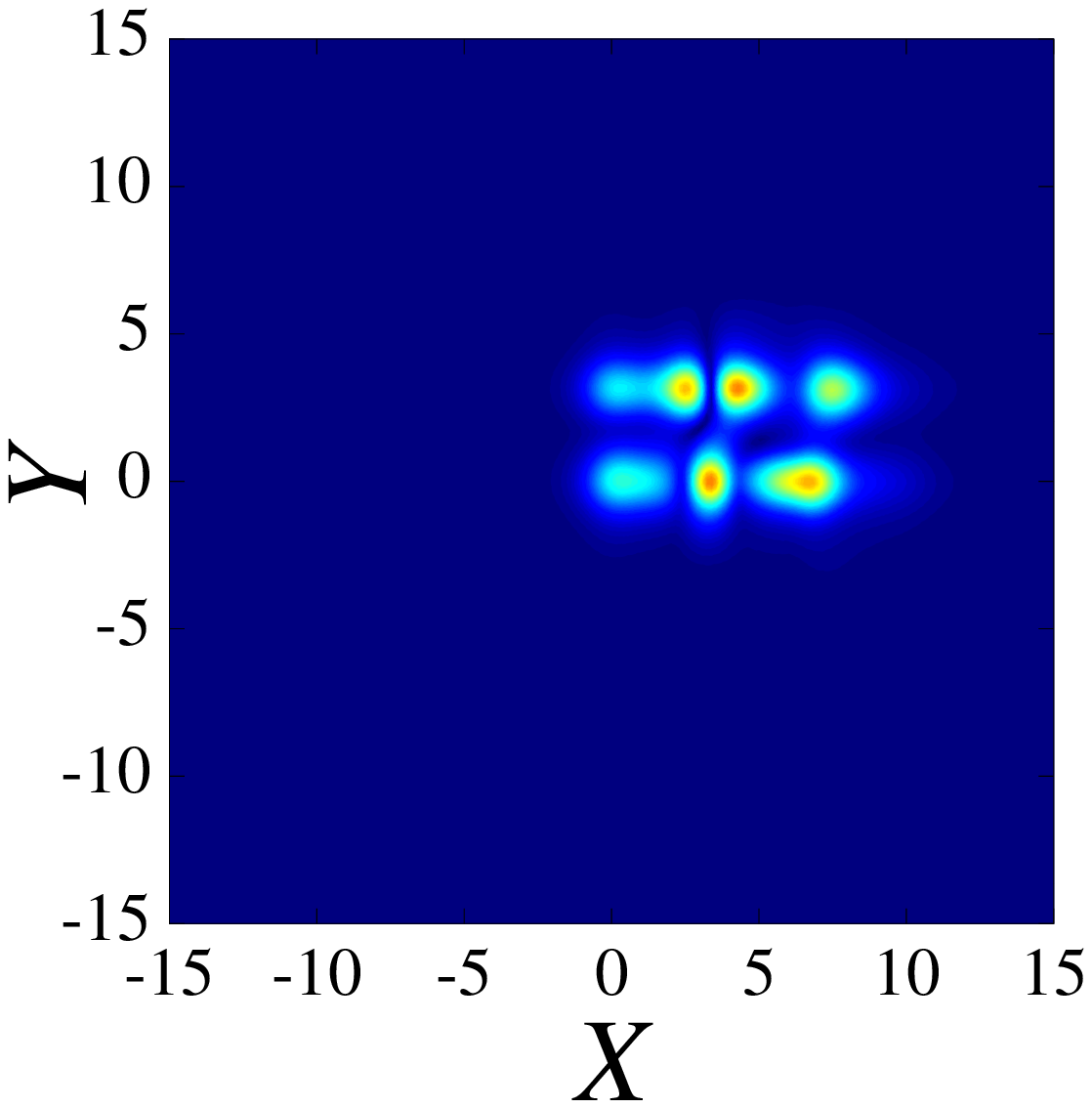}}\vfill
\subfigure[$Z=5.1432$]{%
\includegraphics[width=5cm]{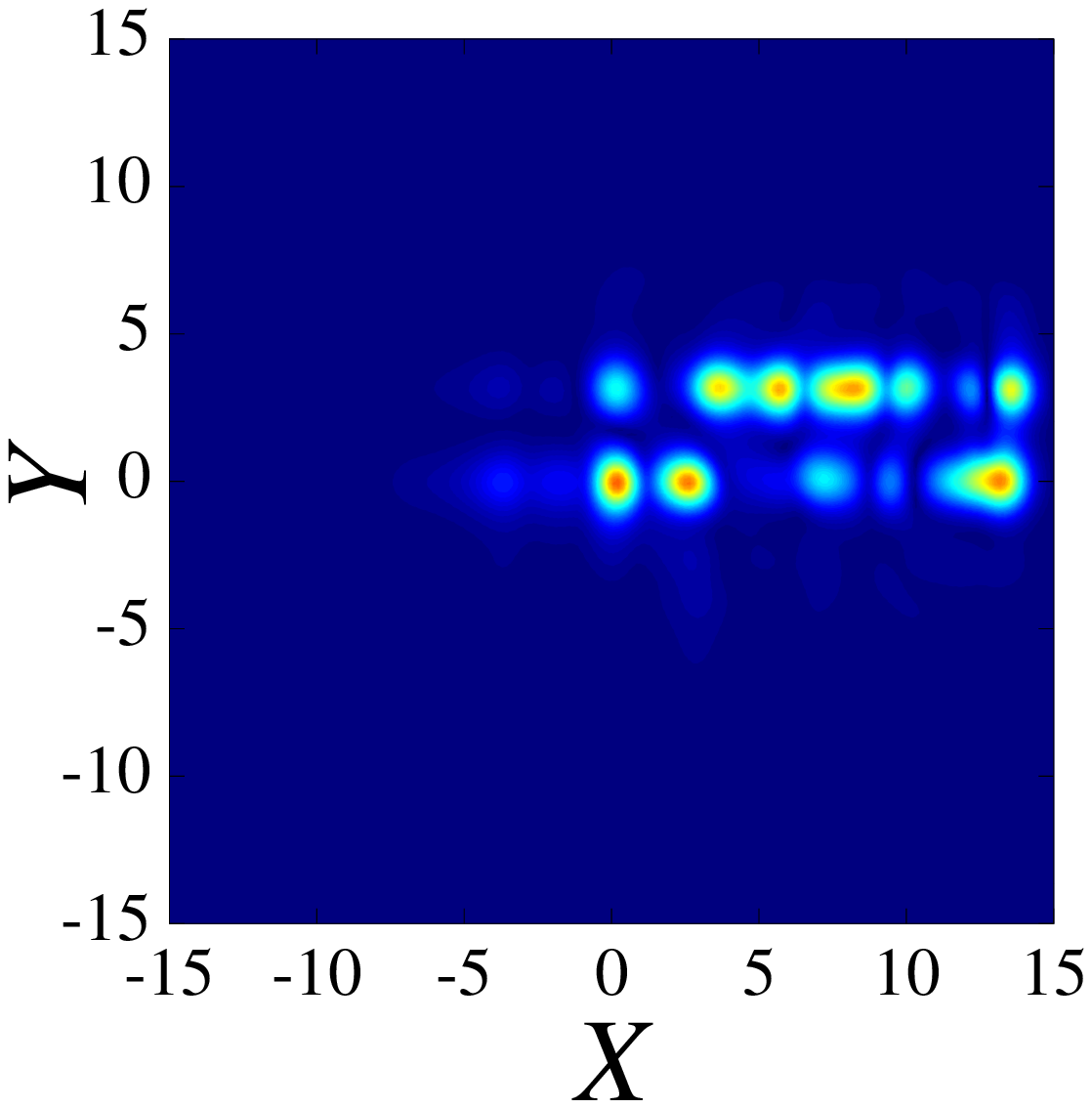}}%
\subfigure[$Z=7.5069$]{%
\includegraphics[width=5cm]{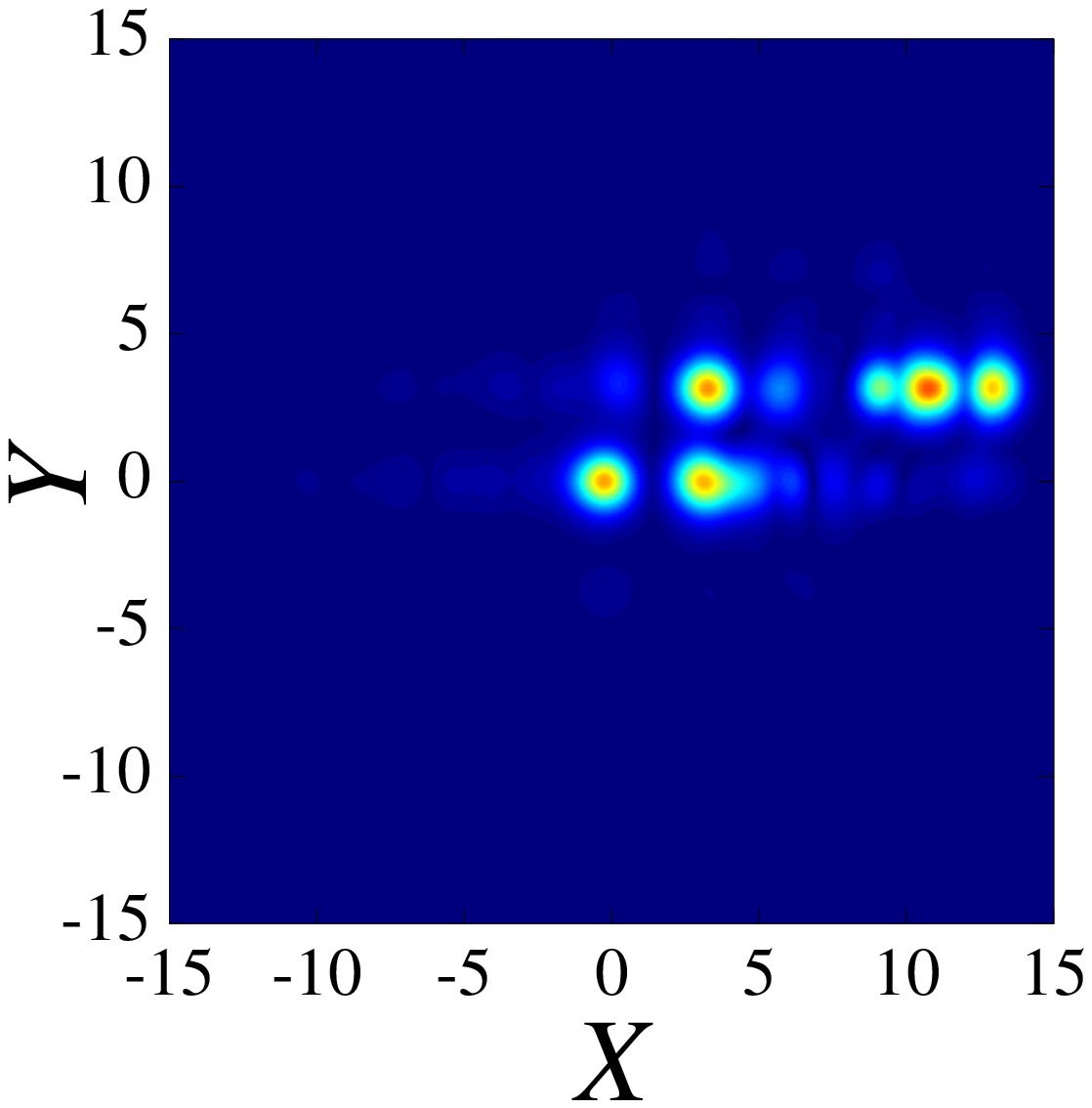}} %
\caption{(Color online) The evolution pattern produced by
horizontally kicking the offsite-centered vortex, for $k_{0}=2$. The
color code is the same as in Fig. 1a.} \label{sqc_k0_2}
\end{figure}

At somewhat higher values of $k_{0}$ (for example, $k_{0}=2.0$), the
original four-soliton set is transformed into a quiescent three-soliton
complex, while an extra dipole and separate free solitons are created and
travel through the lattice, see Fig. \ref{sqc_k0_2}.

\begin{figure}[th]
\centering
\includegraphics[width=10cm]{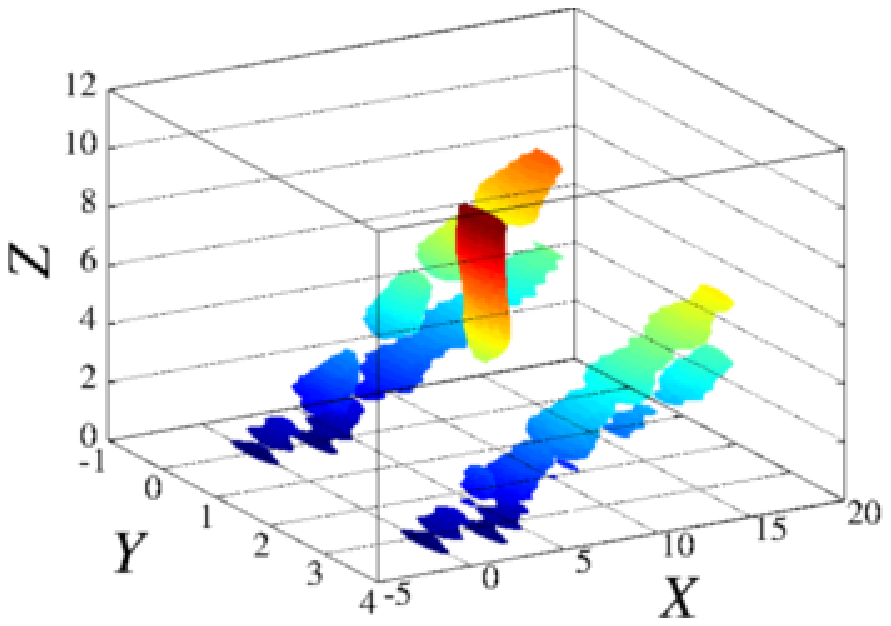}
\caption{(Color online) The three-dimensional rendition of the
evolution of the horizontally kicked offsite-centered vortex for
$k_{0}=2.5$, which is transformed into a stably moving four-soliton
cluster. The chromatic progression indicates the propagation
direction. The vertical rod represents the additional quiescent
fundamental soliton, left in the wake of the moving four-soliton
cluster.} \label{sqc_k0_25_xyz}
\end{figure}

\begin{figure}[th]
\centering
\includegraphics[width=10cm]{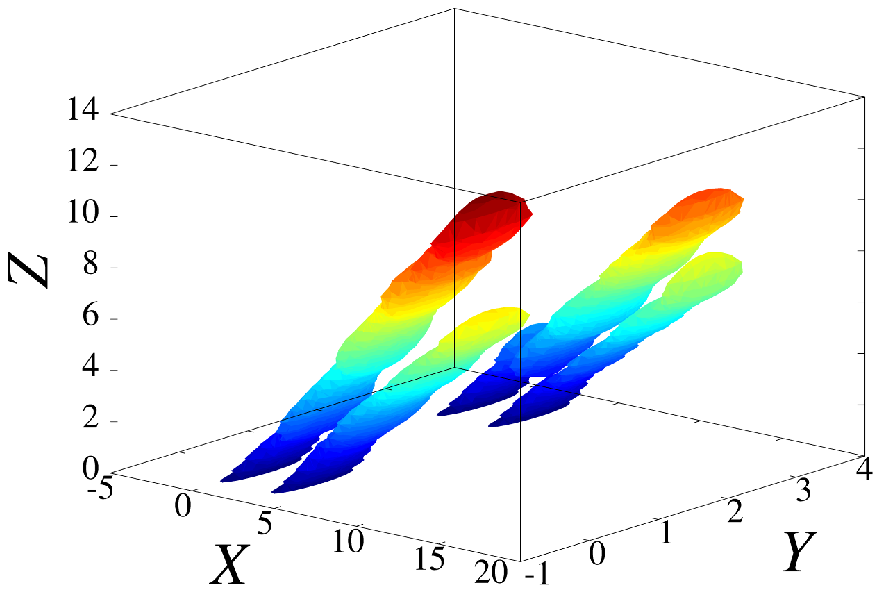}
\caption{(Color online) The same as in Fig. \protect\ref{sqc_k0_25_xyz} but for $%
k_{0}=3.0$. In this case, the unstable vortex is entirely
transformed into the stable moving cluster.} \label{sqc_k0_30_xyz}
\end{figure}

Finally, a still stronger kick applied to the square-shaped vortex
transforms it
into a square-shaped cluster of four solitons moving as a whole, see Figs. %
\ref{sqc_k0_25_xyz} and
\ref{sqc_k0_30_xyz}, which display the result in the 3D form. In the
former case, at $k_0=2.5$, the cluster leaves behind a copy of one
of the original solitons, while at $k_0=3.0$ the moving cluster is
the single emerging mode. Although the clusters are dynamically
stable, they do not carry the vortical phase structure.

\begin{figure}[th]
\begin{center}
\subfigure[$Z=299.885$]{%
\includegraphics[width=5cm]{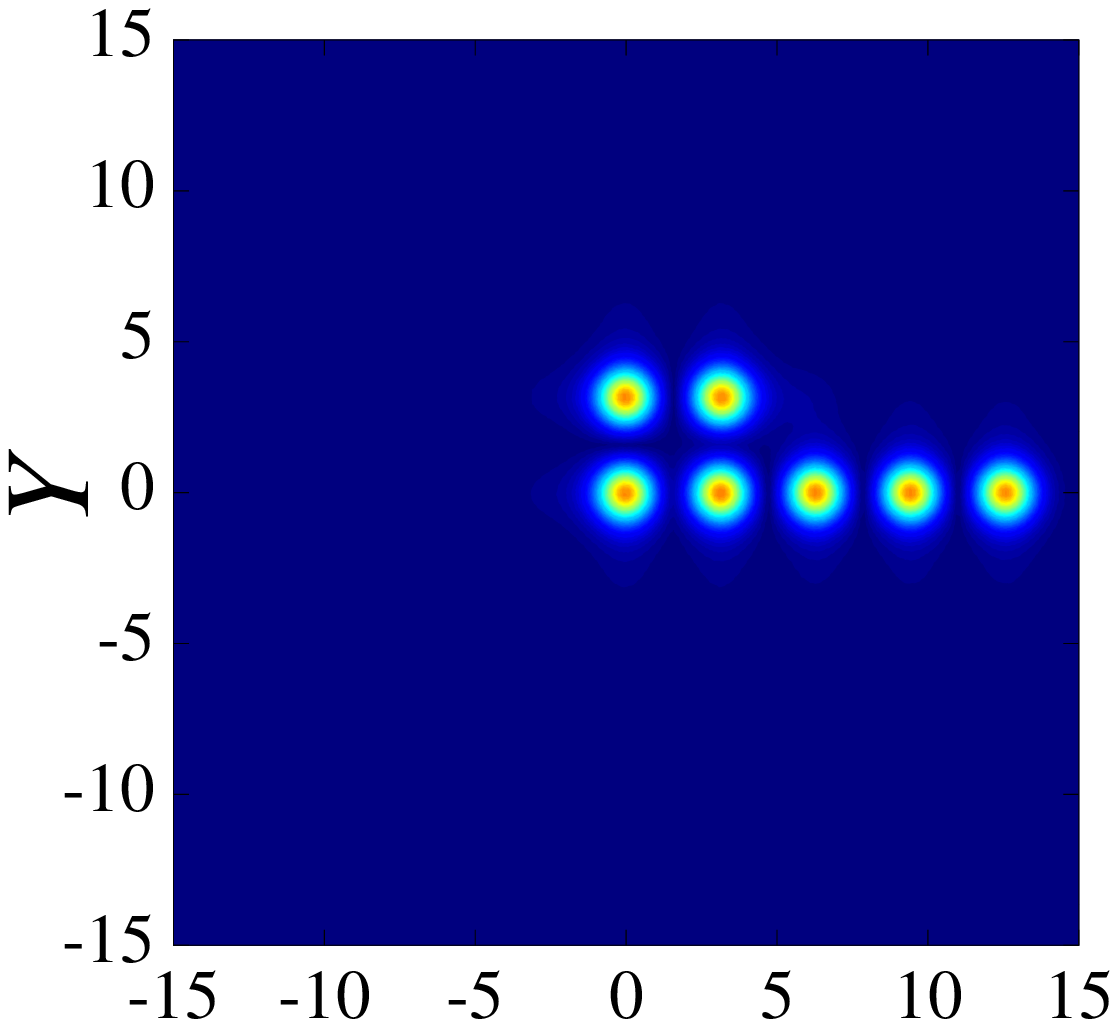}} %
\subfigure[$Z=299.765$]{%
\includegraphics[width=5cm]{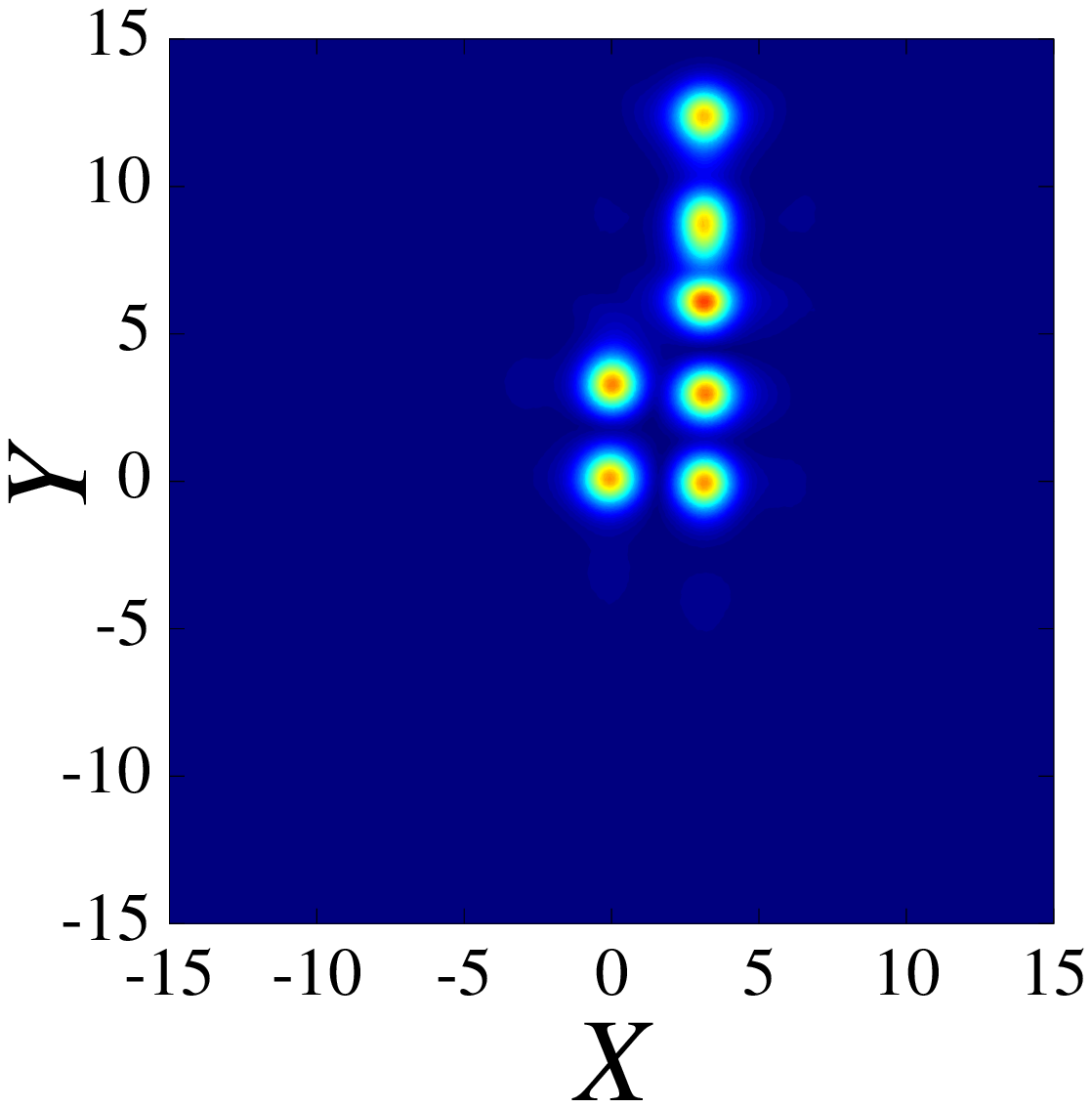}}\vfill
\subfigure[$Z=299.605$]{%
\includegraphics[width=5cm]{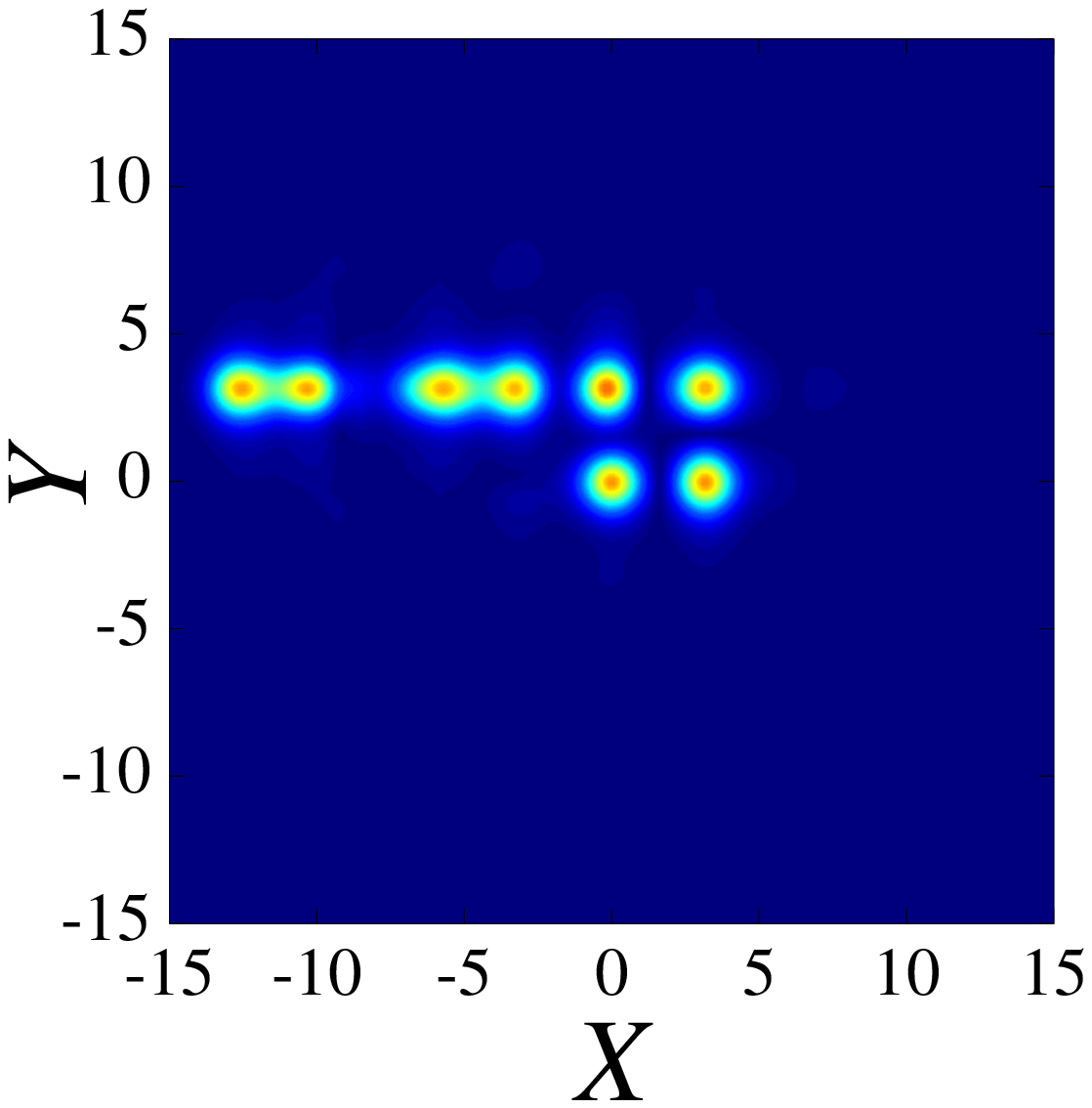}} %
\subfigure[$Z=299.495$]{%
\includegraphics[width=5cm]{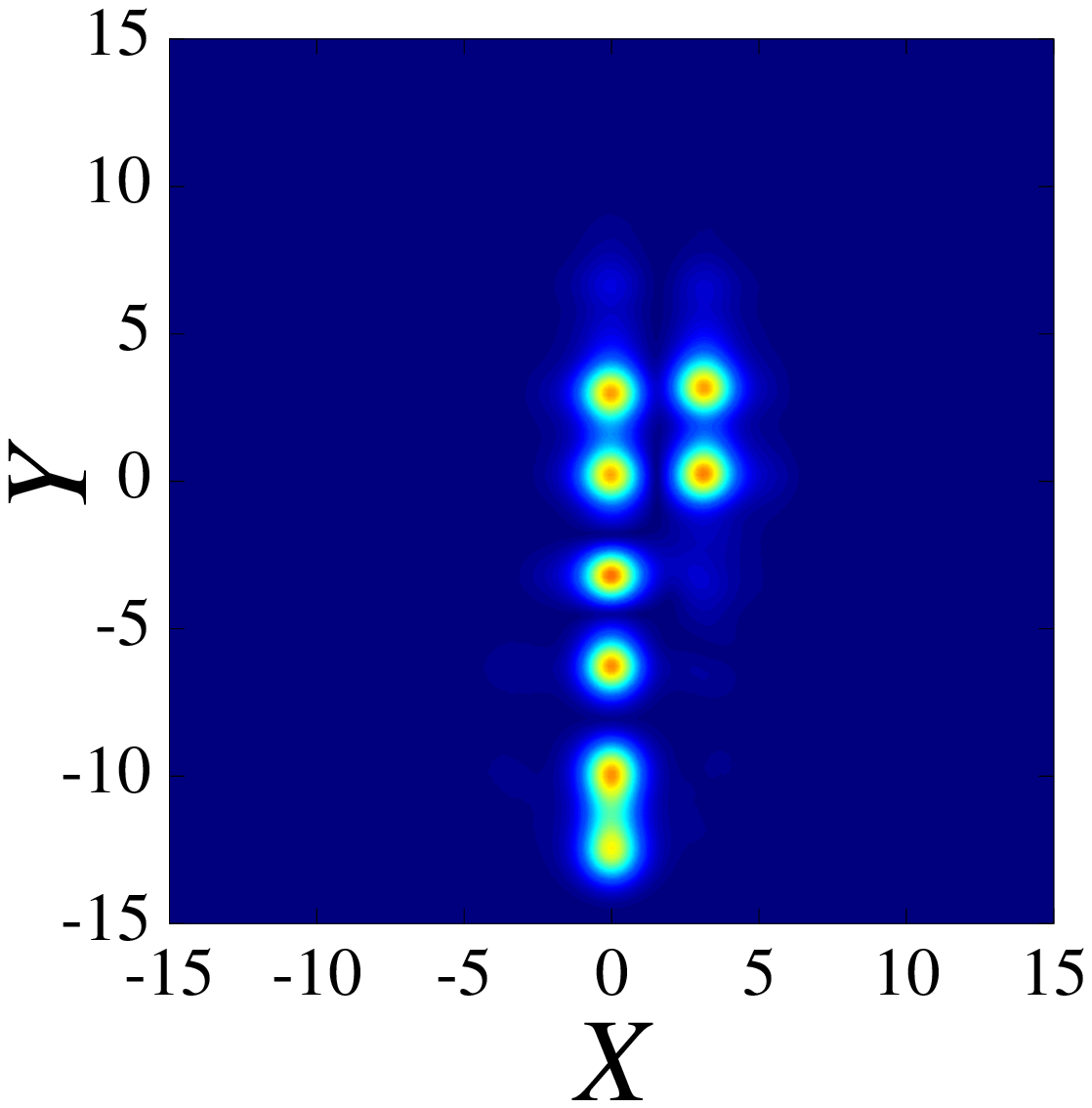}}
\end{center}
\caption{(Color online) The pattern produced by the offsite-centered
vortex kicked with strength $k_{0}=1.5$ in different but actually
equivalent
directions: (a) $\protect\theta ={\protect\pi }/{8}$; (b) $\protect\theta ={5%
\protect\pi }/{8}$; (c) $\protect\theta ={9\protect\pi }/{8}$; (d) $\protect%
\theta ={13\protect\pi }/{8}$. The color code is the same as in Fig.
1a.} \label{sqc_k0_15_theta}
\end{figure}

We have also studied the application of the kick to the
offsite-centered vortex in other directions, i.e., varying angle
$\theta $ in Eq. (\ref{theta}).
First, as seen in Fig. \ref{sqc_k0_15_theta}(a), in the case of $\theta ={%
\pi }/{8}$ and $k_{0}=1.5$, the kick again breaks the symmetry between the
top and bottom rows of the solitons, generating an array of additional
solitons in the bottom horizontal row. Further, to check that the numerical
code is compatible with the global symmetry of the setting, we also
considered equivalent angles, $\theta ={5\pi }/{8}$, ${9\pi }/{8}$ and ${%
13\pi }/{8}$. The results, shown in Fig. \ref{sqc_k0_15_theta}, evidence the
possibility of controlling the direction of the emission of the soliton
array by the direction of the initial kick.

Further, running the computations for varying $\theta $ and moderate values
of $k_{0}$, we have concluded that there is a threshold angle $\varepsilon $%
, so that the emission towards any of the four equivalent directions,
corresponding to directions $\phi =0$, $\pi /2$, $\pi $ or $3\pi /2$, occurs
provided that the orientation of the kick belongs to a certain range around
this direction, \textit{viz}., $\left( \phi -\pi /4+\varepsilon \right)
<\theta <\left( \phi +\pi /4-\varepsilon \right) $, with $\varepsilon =0.059$%
. If the kick's orientation falls into interstices between these ranges,
namely,$[\phi +{\pi }/{4}-\varepsilon ;\phi +{\pi }/{4}+\varepsilon ]$,
solitons arrays are not generated. In the latter case, the square vortex
transforms into a quadrupole.

These results can be explained by noting that the intrinsic phase
circulation in the vortex is directed counterclockwise (from $X$ to $Y$).
Then, as schematically shown (for example) for $\theta =\pi /8$ in Fig. \ref%
{sens_vort}, the superposition of the externally applied kick (the phase
gradient) and the intrinsic phase flow gives rise to the largest local phase
gradient at the position of the bottom right soliton, in the positive
horizontal direction, therefore the array is emitted accordingly.

\begin{figure}[th]
\begin{center}
\includegraphics[width=10cm]{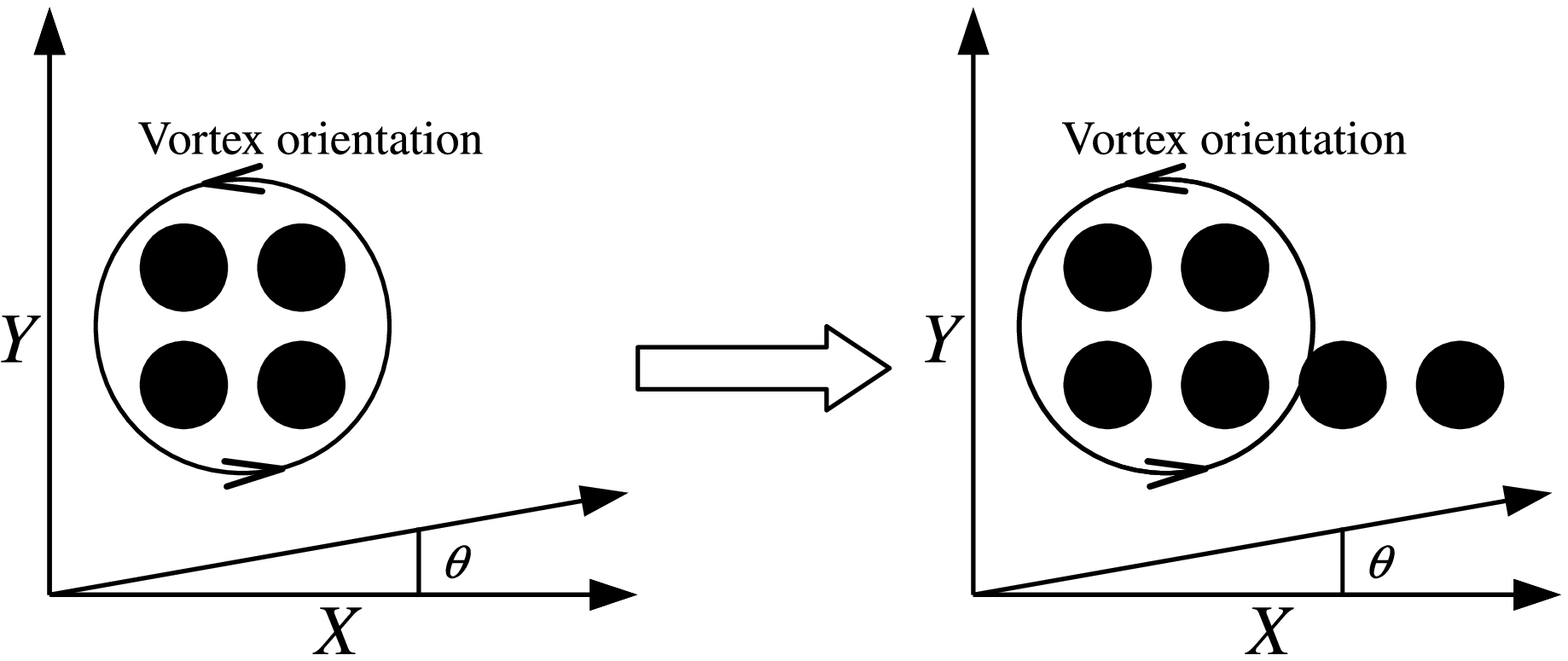}
\end{center}
\caption{The explanation of the direction in which the soliton array
is emitted from the kicked offsite-centered vortex.}
\label{sens_vort}
\end{figure}

It is instructive too to perform the simulations for the vortex with the
opposite vorticity ($-1$ instead of $+1$). In this way, the expected
symmetry reversal has been verified (not shown here in detail): The same
results as above are obtained, with angle $\theta $ replaced by its
counterpart, which is symmetric with respect to the closest coordinate axis.
We note too that identical results were obtained using both periodic and
absorbing boundary conditions.

\section{Conclusions}

The objective of this work is to extend the analysis of the
dynamical pattern-formation scenarios in the CQ-CGL (cubic-quintic
complex Ginzburg-Landau) equation with the 2D cellular potential.
The equation is the model for laser cavities with built-in gratings,
represented by the spatially periodic potential. Recently, the
quasi-1D pattern-formation scenarios, initiated by the moving
fundamental solitons, were studied in this model. Here, we have
systematically analyzed the fully 2D scenarios, produced by kicking
compound modes, \textit{viz}., dipoles, offsite-centered
quadrupoles, and vortices of two different types, onsite- and
offsite-centered ones. The motion of the kicked compound through the
cellular potential leads to the generation of diverse multi-peak
patterns pinned to the lattice, which the moving object leaves in
its wake. In the annular system with periodic boundary conditions,
the persistently traveling dipole hits the pinned pattern from the
opposite direction. In this way, several dynamical regimes are
initiated, including the periodically recurring elastic passage of
the free dipole through the quiescent one, and transient regimes,
which lead, after a few quasi-elastic collisions, to absorption of
one dipole by the other. In the case of vortices, the dependence of
the outcome on the magnitude and direction of the kick was
investigated too. In particular, a noteworthy result is that a
strong kick transforms the original offsite-centered vortex (which
is unstable by itself) into a clean stably moving four-soliton
cluster.

The analysis can be extended by considering two-component systems (which
would take the polarization of light into account), collisions between
independently created moving modes, and the motion of kicked solitons in
inhomogeneous lattices. Eventually, the analysis may be generalized for the
three-dimensional setting, which is not relevant to optics, but may be
realized, in principle, in terms of Bose-Einstein condensates of
polariton-exciton quasiparticles \cite{BEC-exciton}.

\section*{Acknowledgements}

The work of DM was supported in part by a Senior Chair Grant from the R{\'{e}%
}gion Pays de Loire, France. Support from the Romanian Ministry of Education
and Research (Project PN-II-ID-PCE-2011-3-0083) is also acknowledged by this
author.



\begin{thebibliography}{99}
\bibitem{DS} N. Akhmediev and A. Ankiewicz (Eds.), \textit{Dissipative
Solitons}, Lect. Notes Phys. \textbf{661}, Springer, Berlin, 2005; N.
Akhmediev and A. Ankiewicz (Eds.), \textit{Dissipative Solitons: From Optics
to Biology and Medicine}, Lect. Notes Phys. \textbf{751}, Springer, Berlin,
2008.

\bibitem{Rosa} N. N. Rosanov, \textit{Spatial Hysteresis and Optical Patterns%
} (Springer, Berlin, 2002).

\bibitem{lasers} S. Barland, J. R. Tredicce, M. Brambilla, L. A. Lugiato, S.
Balle, M. Giudici, T. Maggipinto, L. Spinelli, G. Tissoni, T. Kn\"{o}dl, M.
Miller, and R. J\"{a}ger, Nature (London) \textbf{419}, 699 (2002); Z.
Bakonyi, D. Michaelis, U. Peschel, G. Onishchukov, and F. Lederer, J. Opt.
Soc. Am. B \textbf{19}, 487 (2002); E. A. Ultanir, G. I. Stegeman, D.
Michaelis, C. H. Lange, and F. Lederer, Phys. Rev. Lett. \textbf{90}, 253903
(2003); P. Mandel and M. Tlidi, J. Opt. B: Quantum Semiclass. Opt. \textbf{6}%
, R60 (2004); N. N. Rosanov, S. V. Fedorov, and A. N. Shatsev, Appl. Phys. B
\textbf{81}, 937 (2005); C. O. Weiss and Ye. Larionova, Rep. Progr. Phys.
\textbf{70}, 255 (2007); N. Veretenov and M. Tlidi, Phys. Rev. A \textbf{80}%
, 023822 (2009); M. Tlidi, A. G. Vladimirov, D. Pieroux, and D. Turaev,
Phys. Rev. Lett. \textbf{103}, 103904 (2009); P. Genevet, S. Barland, M.
Giudici, and J. R. Tredicce, Phys. Rev. Lett. \textbf{104}, 223902 (2010);
P. Grelu and N. Akhmediev, Nature Photonics, \textbf{6}, 84 (2012); J. Jim%
\'{e}nez, Y. Noblet, P. V. Paulau, D. Gomila, and T. Ackemann, J. Opt.
\textbf{15}, 044011 (2013); C. Fernandez-Oto, M. G. Clerc, D. Escaff, and M.
Tlidi, Phys. Rev. Lett. \textbf{110}, 174101 (2013).

\bibitem{plasmonics} N. Lazarides and G. P. Tsironis, Phys. Rev. E \textbf{71%
}, 036614 (2005); Y. M. Liu, G. Bartal, D. A. Genov, and X. Zhang, Phys.
Rev. Lett. \textbf{99}, 153901 (2007); E. Feigenbaum and M. Orenstein, Opt.
Lett. \textbf{32}, 674 (2007); I. R. Gabitov, A. O. Korotkevich, A. I.
Maimistov, and J. B. Mcmahon, Appl. Phys. A \textbf{89}, 277 (2007); A. R.
Davoyan, I. V. Shadrivov, and Y. S. Kivshar, Opt. Exp. \textbf{17}, 21732
(2009); K. Y. Bliokh, Y. P. Bliokh, and A. Ferrando, Phys. Rev. A \textbf{79}%
, 041803 (2009); E. V. Kazantseva and A. I. Maimistov, \textit{ibid}.
\textbf{79}, 033812 (2009); Y.-Y. Lin, R.-K. Lee, and Y. S. Kivshar, Opt.
Lett. \textbf{34}, 2982 (2009); A. Marini and D. V. Skryabin, \textit{ibid}.
\textbf{81}, 033850 (2010); A. Marini, D. V. Skryabin, and B. A. Malomed,
Opt. Exp. \textbf{19}, 6616 (2011).

\bibitem{Petv} V. I. Petviashvili and A. M. Sergeev, Dokl. AN SSSR \textbf{\
276}, 1380 (1984) [Sov. Phys. Doklady \textbf{29}, 493 (1984)].

\bibitem{AK} I. S. Aranson and L. Kramer, Rev. Mod. Phys. \textbf{74}, 99
(2002); B. A. Malomed, in \textit{Encyclopedia of Nonlinear Science}, p.
157. A. Scott (Ed.), Routledge, New York, 2005.

\bibitem{BEC} J. Anglin, Phys. Rev. Lett. \textbf{79}, 6 (1997); F. T.
Arecchi, J. Bragard, and L. M. Castellano, Opt. Commun.
\textbf{179}, 149 (2000); J. Keeling and N. G. Berloff, Phys. Rev.
Lett. \textbf{100}, 250401 (2008); B. A. Malomed, O. Dzyapko, V. E.
Demidov, and S. O. Demokritov, Phys. Rev. B \textbf{81}, 024418
(2010).

\bibitem{MCCROSS} M. C. Cross and P. C. Hohenberg, Rev. Mod. Phys. \textbf{65%
}, 851 (1993).

\bibitem{supercond} K.-H. Hoffmann and Q. Tang, \textit{Ginzburg-Landau
Phase Transition Theory and Superconductivity} (Birkhauser Verlag: Basel,
2001).

\bibitem{PhysicaD} B. A. Malomed, Physica D \textbf{29}, 155 (1987).

\bibitem{Boris} O. Thual and S. Fauve, J. Phys. (Paris) \textbf{49}, 1829
(1988); S. Fauve and O. Thual, Phys. Rev. Lett. \textbf{64}, 282 (1990); W.
van Saarloos and P. C. Hohenberg, Phys. Rev. Lett. \textbf{64}, 749 (1990);
V. Hakim, P. Jakobsen, and Y. Pomeau, Europhys. Lett. \textbf{11}, 19
(1990); B. A. Malomed and A. A. Nepomnyashchy, Phys. Rev. A \textbf{42},
6009 (1990); P. Marcq, H. Chat\'{e}, R. Conte, Physica D \textbf{73}, 305
(1994); N. Akhmediev and V. V. Afanasjev, Phys. Rev. Lett. \textbf{75}, 2320
(1995); H. R. Brand and R. J. Deissler, Phys. Rev. Lett. \textbf{63}, 2801
(1989); V. V. Afanasjev, N. Akhmediev, and J. M. Soto-Crespo, Phys. Rev. E
\textbf{53}, 1931 (1996); J. M. Soto-Crespo, N. Akhmediev, and A. Ankiewicz,
Phys. Rev. Lett. \textbf{85} , 2937 (2000); H. Leblond, A. Komarov, M.
Salhi, A. Haboucha, and F. Sanchez, J. Opt. A: Pure Appl. Opt. \textbf{8},
319 (2006); W. H. Renninger, A. Chong, and F. W. Wise, Phys. Rev. A \textbf{%
77}, 023814 (2008); J. M. Soto-Crespo, N. Akhmediev, C. Mejia-Cortes, and N.
Devine, Opt. Express \textbf{17}, 4236 (2009); D. Mihalache, Proc. Romanian
Acad. A \textbf{11}, 142 (2010); Y. J. He, B. A. Malomed, D. Mihalache, F.
W. Ye, and B. B. Hu, J. Opt. Soc. Am. B \textbf{27}, 1266 (2010); D.
Mihalache, Rom. Rep. Phys. \textbf{63}, 325 (2011); C. Mejia-Cortes, J. M.
Soto-Crespo, R. A. Vicencio, and M. I. Molina, Phys. Rev. A \textbf{83},
043837 (2011); D. Mihalache, Rom. J. Phys. \textbf{57}, 352 (2012); O. V.
Borovkova, V. E. Lobanov, Y. V. Kartashov, and L. Torner, Phys. Rev. A
\textbf{85}, 023814 (2012).

\bibitem{Mihalache} L.-C. Crasovan, B. A. Malomed, and D. Mihalache, Phys.
Rev. E \textbf{63}, 016605 (2001); Phys. Lett. A \textbf{289}, 59 (2001); D.
Mihalache, D. Mazilu, F. Lederer, Y. V. Kartashov, L.-C. Crasovan, L.
Torner, and B. A. Malomed, Phys. Rev. Lett. \textbf{97}, 073904 (2006); D.
Mihalache, D. Mazilu, F. Lederer, H. Leblond, and B. A. Malomed, Phys. Rev.
A \textbf{76}, 045803 (2007); \textit{ibid}. \textbf{75}, 033811 (2007); D.
Mihalache and D. Mazilu, Rom. Rep. Phys. \textbf{60}, 749 (2008).

\bibitem{Tlidi} M. Tlidi, M. Haelterman, and P. Mandel, Europhys. Lett.
\textbf{42}, 505 (1998); M. Tlidi and P. Mandel, Phys. Rev. Lett. \textbf{83}%
, 4995 (1999); M. Tlidi, J. Opt. B: Quantum Semiclass. Opt. \textbf{2}, 438
(2000).

\bibitem{HS} H. Sakaguchi, Physica D \textbf{210}, 138 (2005).

\bibitem{Vladimir} V. Skarka and N. B. Aleksi{\'{c}}, Phys. Rev. Lett.
\textbf{96}, 013903 (2006); N. B. Aleksi{\'{c}}, V. Skarka, D. V. Timotijevi{%
\'{c}}, and D. Gauthier, Phys. Rev. A \textbf{75}, 061802 (2007); V. Skarka,
D. V. Timotijevi{\'{c}}, and N. B. Aleksi{\'{c}}, J. Opt. A: Pure Appl. Opt.
\textbf{10}, 075102 (2008); V. Skarka, N. B. Aleksi{\'{c}}, H. Leblond, B.
A. Malomed, and D. Mihalache, Phys. Rev. Lett. \textbf{105}, 213901 (2010).

\bibitem{Jena} A. Szameit, J. Burghoff, T. Pertsch, S. Nolte, and A. T\"{u}%
nnermann, Opt. Exp. \textbf{14}, 6055 (2006).

\bibitem{Moti-general} J. W. Fleischer, M. Segev, N. K. Efremidis, and D. N.
Christodoulides, Nature \textbf{422}, 147 (2003).

\bibitem{leblond1} H. Leblond, B. A. Malomed, and D. Mihalache, Phys. Rev. A
\textbf{80}, 033835 (2009).

\bibitem{tlidi94} M. Tlidi, Paul Mandel, and R. Lefever, Phys. Rev. Lett. \textbf{73}, 640 (1994).


\bibitem{Firth} W. J. Firth and A. J. Scroggie, Phys. Rev. Lett. \textbf{76}%
, 1623 (1996).

\bibitem{advances} M. Brambilla, A. Gatti and L. A. Lugiato, Adv. At. Molec.
Opt. Phys. \textbf{40}, 229 (1998).

\bibitem{Fedorov} S. V. Fedorov, A. G. Vladimirov, G. V. Khodova, N. N.
Rosanov, Phys. Rev. E \textbf{61}, 5814 (2000).

\bibitem{trapping_potentials_2D} D. Mihalache, D. Mazilu, V. Skarka, B. A.
Malomed, H. Leblond, N. B. Aleksi{\'{c}}, and F. Lederer, Phys. Rev. A
\textbf{82}, 023813 (2010).

\bibitem{trapping_potentials_3D} D. Mihalache, D. Mazilu, F. Lederer, H.
Leblond, and B. A. Malomed, Phys. Rev. A \textbf{81}, 025801 (2010).

\bibitem{cgl_mov} V. Besse, H. Leblond, D. Mihalache, and B. A. Malomed,
Phys. Rev. E \textbf{87}, 012916 (2013).




\bibitem{BBB} B. B. Baizakov, B. A. Malomed, and M. Salerno, Europhys. Lett.
\textbf{63}, 642 (2003); J. Yang and Z. H. Musslimani, Opt. Lett. \textbf{28}%
, 2094 (2003).

\bibitem{Thawatchai} T. Mayteevarunyoo, B. A. Malomed, B. B. Baizakov, and
M. Salerno, Physica D \textbf{238}, 1439 (2009).

\bibitem{Yang} J. Yang, \textit{Nonlinear Waves in Integrable and
Nonintegrable Systems} (SIAM:\ Philadelphia, 2010).

\bibitem{BEC-exciton} H. Deng, H. Haug, and Y. Yamamoto, Rev. Mod. Phys.
\textbf{82}, 1489 (2010); B. Deveaud-Pl\'{e}dran, J. Opt. Soc. Am. B \textbf{29},
A138 (2012); N. G. Berloff and J. Keeling, (ed. A. Bramati and M. Modugno),
\textit{Universality in modelling non-equilibrium
pattern formation in polariton condensates}, Physics of Quantum
Fluids, Springer, pp. 19-38 (2013)
\end{thebibliography}
\end{document}